\documentclass[fleqn,usenatbib]{mnras}

\usepackage{newtxtext,newtxmath}
\usepackage[T1]{fontenc}

\DeclareRobustCommand{\VAN}[3]{#2}
\let\VANthebibliography\thebibliography
\def\thebibliography{\DeclareRobustCommand{\VAN}[3]{##3}\VANthebibliography}

\usepackage{graphicx}	% Including figure files
\usepackage{amsmath}	% Advanced maths commands

\usepackage{subcaption} % Allowing for subplots
\usepackage{multirow} % Multiple rows occupied within a table
\usepackage[normalem]{ulem}% to use sout
\usepackage{braket}    % Use Bra Ket notation

\newcommand{\kms}{$\mathrm{km\,s^{-1}}$} % km per second
\providecommand{\noopsort}[1]{} % For the Team COMPAS reference

\title[Binary formation in star clusters]{Black hole binary mergers in dense star clusters: the importance of primordial binaries}

\author[J Barber et al.]{
Jordan Barber$^{1}$\thanks{E-mail: barberj2@cardiff.ac.uk},
Debatri Chattopadhyay,$^{1}$,
Fabio Antonini$^{1}$
\\
$^{1}$Gravity Exploration Institute, School of Physics and Astronomy, Cardiff University, Cardiff, CF24 3AA, UK
}

\date{Accepted XXX. Received YYY; in original form ZZZ}

\pubyear{2023}

\begin{document}
\label{firstpage}
\pagerange{\pageref{firstpage}--\pageref{lastpage}}
\maketitle

\begin{abstract}
Dense stellar clusters are expected to house the ideal conditions for binary black hole (BBH) formation, both through binary stellar evolution  and through  dynamical encounters. We use theoretical arguments as well as $N$-body simulations to make predictions for the evolution of BBHs formed through stellar evolution inside clusters from the cluster birth (which we term \textbf{primordial binaries}), and for the sub-population of merging BBHs. We identify three key populations: 
(i) BBHs that form in the cluster, and merge before experiencing any {\it strong} dynamical interaction;
(ii) binaries that are ejected from the cluster after only one dynamical interaction; and, (iii) BBHs that experience more than one strong interaction inside the cluster.
We find that population (i) and (ii) are the dominant source of all BBH mergers formed in clusters with escape velocity $v_{\mathrm{esc}}\leq 30$ \kms. At higher escape velocities, dynamics are predicted to play a major role both for the formation and subsequent evolution of BBHs. Finally, we argue that for sub-Solar metallicity clusters with $v_{\mathrm{esc}}\lesssim100$ \kms, the dominant form of interaction experienced by primordial BBHs (BBHs formed from primordial binaries) within the cluster is with other BBHs. The complexity of these binary-binary interactions will complicate the future evolution of the BBH and influence the total number of mergers produced. 
%determined by the binary separation and the assumed escape velocity of a cluster. These populations differ in the number of expected interactions they experience within the cluster before they merge or are ejected. We then utilise the N-body code PeTar to model the dynamics between stellar systems within the star cluster and identify these same populations of BBHs from the primordial binaries. We find that the dominant source of BBH mergers in clusters with $v_{\mathrm{esc}}\leq 30$ \kms are binaries that experience no dynamical interactions before merging. In addition, we show that at most the dynamically formed BBHs can only produce twice the number of BBH mergers than the primordial binaries. Finally, we use some theoretical arguments to show that for sub-Solar metallicity clusters with $v_{\mathrm{esc}}\lesssim100$ \kms the dominant form of interaction experienced by the BBHs within the cluster is with other BBHs. The complexity of these binary-binary interactions will complicate the future evolution of these BBHs and will likely disrupt those that remain with the cluster, thus influencing the total number of mergers from these clusters. 
\end{abstract}


%%%%%%%%%%%%%%%%%%%%%%%%%%%%%%%%%%%%%%%%%%%%%%%%%%

%%%%%%%%%%%%%%%%% BODY OF PAPER %%%%%%%%%%%%%%%%%%

\section{Introduction}
In 2015 the Laser Interferometer Gravitational-Wave Observatory (LIGO) observed, for the first time, a gravitational wave (GW) from the coalescence of a binary black hole (BBH) system \citep{abbott_observation_2016}. This detection marked a breakthrough in the field of GW astronomy, and now nearly 8 years later we have close to 100 confirmed observations of GWs from compact object mergers, the majority of which are from BBH systems \citep{abbott_binary_2016, abbott_gwtc-2_2021, abbott_population_2023, the_ligo_scientific_collaboration_gwtc-3_2021}. With the advent of the fourth LIGO observing run this year, we should expect these numbers to at least double. However, the astrophysical origin of these systems is still up for debate. 

Presently, it is believed that these BBH mergers are coming from two main sources; isolated evolution of massive stellar binaries in the galactic field \citep[e.g.,][]{belczynski_effect_2010, mandel_merging_2016, michaely_gravitational-wave_2019, mapelli_binary_2020,belczynski_evolutionary_2020, broekgaarden_impact_2021}, and formation through dynamical encounters within the core of dense stellar clusters \citep[e.g.,][]{miller_mergers_2009, antonini_merging_2016, rodriguez_binary_2016, banerjee_stellar-mass_2017, banerjee_stellar-mass_2018, dicarlo_merging_2019, hong_binary_2018, mapelli_cosmic_2022, chattopadhyay_dynamical_2022, fragione_merger_2022, torniamenti_dynamics_2022, banerjee_binary_2022, arca_sedda_dragon-ii_2023,arca_sedda_dragon-ii_2023-1}. 

The galactic field mechanism describes an isolated binary formation channel where the binaries experience little to no interactions with other stars. Here we have co-evolving massive stars in a binary, which likely undergo some form of common envelope evolution \citep{paczynski_common_1976, ivanova_common_2013} in order to shrink the orbit. Once the separation is sufficiently small, and provided the binary has not been disrupted by either of the binary components going supernovae (SN), the binary evolution becomes dominated by the GW radiation which drives the binary to merge \citep[e.g.,][]{tutukov_evolution_1973, hurley_evolution_2002, dominik_double_2012, de_mink_merger_2015, mandel_merging_2016, spera_merging_2019, farmer_constraints_2020,  mapelli_binary_2020, costa_formation_2021, qin_merging_2023}. On the other hand, the dynamical channel forms BBHs through dynamical encounters between both BHs and BH progenitor stars in the cores of stellar clusters, such as nuclear clusters, globular clusters, open clusters and young clusters. When a binary forms, it experiences further encounters and through these many-body interactions, the binary hardens and is driven to the regime where GW emission dominates the further evolution until the merger
\citep{quinlan_dynamical_1996, banerjee_stellar-mass_2010, ziosi_dynamics_2014, mapelli_massive_2016, antonini_merging_2016, antognini_dynamical_2016, dicarlo_merging_2019, antonini_black_2019, anagnostou_dynamically_2020, arca_sedda_isolated_2023}.

The distinction between these two formation channels can become more complex when considering stellar clusters. Observations of young open clusters demonstrate a high fraction of binaries ($>70\%$) amongst massive O/B type stars \citep{sana_binary_2012}. Previous work has also suggested that the fraction of these stars in higher multiplicity systems (triples, quadruples) is even larger \citep{moe_mind_2017}. Massive stars can evolve to form BHs on a timescale of $10^{6}$ years, thus having the potential to form BBHs in the early stages of the host cluster's evolution. These stellar binaries which formed with the birth of the cluster are termed "\textit{primordial}" binaries; since they are not assembled by dynamical interactions but by stellar processes, instead. 
Subsequent to their formation, primordial binaries can  experience dynamical encounters with other cluster members that can change their orbital properties \citep[e.g.,][]{samsing_formation_2014}. This can happen either before or after the binary has evolved to form a BBH.
Therefore, we will have a combination of "isolated" binary evolution, which mostly sets the initial properties of the binary, and dynamics, which can alter these properties before a BBH merger is produced. This provides a blend of both formation channels.

In this work, we will focus on  BBH mergers formed from primordial binaries in dense stellar clusters. We investigate how they shape the BBH population and consider their contribution to the merging population. Thus far, they have been shown to be a main source of BBHs mergers in low mass clusters ($\sim500-800 \ M_{\sun}$) \citep{torniamenti_dynamics_2022}, and at sub-Solar metallicity young clusters \citep{di_carlo_binary_2020}. High primordial binary fraction amongst massive stars can also lead to massive BH formation up to $M=300 \, M_{\sun}$ which pushes into the intermediate mass BH range \citep{gonzalez_intermediate-mass_2021, gonzalez_intermediate-mass_2022}. It has also been shown that the existence of primordial binaries halts the core collapse of a cluster much sooner than when they are not included in the model
\citep{trenti_star_2007, pavlik_evolution_2021}. Although there can be primordial binaries across a range of initial stellar masses, \citet{wang_impact_2021} showed that low-mass binaries  have almost no influence on the secular evolution of the cluster until the BH population has been depleted. As such, it is not necessary to consider the effect of the lower mass binaries on the cluster evolution until after the BHs have been removed from the cluster. Thus, in this paper we ignore any primordial binaries in the smaller mass range, and only consider those binaries that can form a BBH. 

Previous studies have shown the importance of BHs on the long-term evolution of a stellar cluster \citep{spitzer_dynamical_1987, binney_galactic_1987, wang_survival_2020, antonini_merger_2020}. Of particular importance is the existence of BBHs which provide a crucial source of energy to the cluster through interactions with surrounding BHs. These encounters \textit{harden} the BBHs and transfer energy to the cluster which prevents complete collapse of the core. Since massive primordial binaries would provide a source of many BBHs within a cluster, the fraction of primordial binaries within a cluster is an important parameter to consider. Currently, it is not clear if the fraction we see within young stellar clusters can be applied across all clusters, so it is beneficial to consider a range of initial binary fractions, in particular looking at the two extreme ends (100\% and 0\%).  In this work, we assume that there is a 100\% binary fraction only amongst the massive stars.
%this is because we are focusing on the evolution of these massive stars to form BBHs. In addition, it has been shown in \citet{wang_impact_2021} that smaller mass binaries are not important in the cluster evolution until the supply of BHs within the cluster has been depleted. 

In Section~\ref{sec:isolated} we look at the BBH distributions formed through purely binary stellar evolution. Section~\ref{sec:retain} then assumes properties of a simplistic cluster model, and attempts to make estimates of the expected BH and BBH retention fraction in stellar clusters. We then further investigate the BBH population retained by clusters in Section~\ref{sec:BBHPops}, making estimates for the sub-population of merging BBHs. This leads to a discussion on the importance of binaries for dense stellar clusters in Section~\ref{sec:importance}. Finally, Section~\ref{sec:final} summarises our findings and provides some discussion on the results.  

\section{Isolated Binary Population Study}
\label{sec:isolated}
We use the fast binary population synthesis code COMPAS \citep{riley_rapid_2022, stevenson_formation_2017} to investigate the properties of the  black hole and  binary black hole population due to stellar evolution for different stellar metallicities. It should be noted that this is only a simplistic model as it does not take into account  the dynamical interaction of binaries with the rest of the cluster, nor does it consider the evolution of the cluster itself. These are discussed in later sections.
%Moreover, we do not include a metallicity distribution within a given cluster,  or correct for the bi-modality in the globular cluster metallicity distribution \citep{harris_globular_2006}. 
Comparing the BH natal kicks against a range of escape velocities ($1$ \kms to $2.5\times10^{3}$ \kms)  we start by investigating here the expected retention fraction of BHs and BBHs within different clusters, owing just to stellar evolution. %The effect of dynamical encounters that occur before the formation of the BBH, and we consider their effect in a later section.
% In addition we investigate the importance of these primordial binaries in long term evolution of the cluster and formation of BBH mergers.

We consider a wide range of cluster escape velocities, which cover open clusters, globular clusters, as well as nuclear clusters \citep{antonini_merging_2016, harris_globular_2006}. It is important to note that, realistically, the escape velocity is an evolving quantity dependent both on the age of the cluster and the position within the cluster. For the work in this paper, we are only concerned with the initial stellar evolution forming BBHs and then the immediate consequences on the BBH populations. Thus it is suitable to only consider the initial escape velocity of the cluster and assume that it is not time-dependent during this early stage. In terms of position, all escape velocities are treated as the cluster central escape velocity going to infinity unless stated otherwise.

\subsection{Initial Conditions}
\label{sec:init}

We run thee models setting the metallicity to $Z=0.01$, $Z=0.001$ and $Z=0.0001$ respectively. Throughout the paper we will refer to the $Z=0.01$ model as an approximately "Solar" metallicity model while the other two models are considered "sub-Solar" metallicity models. Each model simulates $10^{5}$ binaries, evolving the stars from the zero-age main sequence (ZAMS) until compact object (CO) formation, or until a Hubble time has passed. The evolution model is broadly similar to the widely used BSE population synthesis code \citep{hurley_comprehensive_2000, hurley_evolution_2002}. We sample the primary stellar mass  from a Kroupa IMF \citep{kroupa_variation_2001} between $20$ $M_{\sun}$ to\ $150$ $M_{\odot}$ since this lower limit of $20\ M_\odot$ is approximately the minimum stellar mass required to form a BH. The upper limit was left as the default value used by COMPAS.

Once $10^{5}$ primary masses have been assigned, the mass of the secondary  star is drawn from a uniform mass ratio ($q$) distribution in the range $0.01$ to $1$ as found by \citet{sana_binary_2012}. Selecting the secondaries in this manner allows the possibility for a secondary mass less than the minimum primary mass of $20$ $M_{\sun}$, in fact, $\approx61\%$ of all systems have an initial secondary mass smaller than $20$ $M_{\odot}$ across all three models. In these systems, the secondary star is  unlikely to form a BH at the end of its evolution and so they only contribute a single BH to the population. 

With the binary masses chosen, the initial orbital period is drawn from a \citet{sana_binary_2012} distribution, 
\begin{equation}
    f_{\mathrm{p}}(\log_{\mathrm{10}}P)=0.23\times(\log_{\mathrm{10}}P)^{-0.55}, 
\end{equation}
with the minimum period $\log(P_{\mathrm{min}})=0.15$ and the maximum period set such that the distribution is normalised to 1 \citep{oh_dependency_2015}. The binary eccentricity is also drawn from the \citet{sana_binary_2012} distribution, 
\begin{equation}
    f_{\mathrm{p}}(e)\propto e^{-0.45},
\end{equation}
between zero and one. We assume the default COMPAS prescriptions for mass transfer, common envelope evolution (CE) and RLOF; which are detailed in Table~\ref{tab:initCOMPAS}.  
Finally, natal kicks of compact objects formed through core-collapse supernovae (CCSNe) are drawn from a Maxwellian distribution with $\sigma=265$ \kms \citep{hobbs_statistical_2005}, and electron capture supernovae (ECSNe) and ultra stripped supernovae (USSNe) have a Maxwellian $\sigma=30$\,\kms \citep{pfahl_population_2002, podsiadlowski_effects_2004}. We assume a fallback kick prescription for the BH natal kicks \citep{fryer_mass_1999}, whereby we first calculate the  neutron star (NS) kick from the Maxwellian distribution, and then scale this by the fraction of mass that falls onto the proto-compact-object (the fallback fraction) $f_{\mathrm{b}}$. This gives a final BH kick as
\begin{equation}
    v_{\mathrm{BH}}=v_{\mathrm{NS}}(1-f_{\mathrm{b}}),
    \label{eq:fallback}
\end{equation}
Where $v_{\mathrm{NS}}$ is the drawn natal kick for a NS. Since the timescale of a typical SN is much shorter than the evolutionary timescales used in COMPAS; the SN are treated as instantaneous events which affect the binary orbital parameters \citep{riley_rapid_2022}. 
 The resulting binary parameters are calculated following Appendix B of \citet{pfahl_comprehensive_2002} which accounts for the natal kick, instantaneous mass loss, interaction between the SN blast wave and the binary companion, and finally any change to the binary COM velocity \citep{blaauw_origin_1961, hills_effects_1983, brandt_effects_1995, kalogera_orbital_1996, tauris_runaway_1998, hurley_evolution_2002}. As described in \citet{pfahl_comprehensive_2002}, the binary is only flagged as gravitationally unbound if the final eccentricity following the SN is greater than 1.

\begin{table*}
    \caption{Perscriptions assumed for mass transfer and common envelope evolution of the binary. Note that most of these are the default prescription for COMPAS; the full list of default settings is detailed in Table 1 of \citet{riley_rapid_2022}.}
    \label{tab:initCOMPAS}   
    \begin{tabular}{lclclc}
        \hline
        Description & Value/Range & Fiducial parameter settings \\
        \hline
        Mass transfer stability criteria & $\zeta$-prescription & See \citet{vigna-gomez_formation_2018} and references therein \\
        Mass transfer accretion rate & Thermal timescale & Limited by thermal timescale for stars \citet{vigna-gomez_formation_2018, vinciguerra_be_2020} \\
         & Eddington-limited & Accretion rate is Eddington-limited for compact objects \\
         Non-conservative mass loss & Isotropic re-emission & \citet{massevitch_evolution_1975, bhattacharya_formation_1991} \\
          & & \citet{soberman_stability_1997, tauris_formation_2006} \\
          Case BB mass transfer stability & Always stable & Based on \citet{tauris_ultra-stripped_2015, tauris_formation_2017, vigna-gomez_formation_2018} \\
          Circularisation at the onset of RLOF & On & Instantly circularise to periapsis (see Section 4.2 of \citet{riley_rapid_2022}\\
          CE prescription & $\alpha - \lambda$ & Based on \citet{webbink_double_1984, de_kool_common_1990} \\
          CE efficiency $\alpha$-parameter & 3.0 & See section 4.2.4 of \citet{riley_rapid_2022} \\
          CE $\lambda$-parameter & $\lambda_{\mathrm{fixed}}=0.5$ \\
        \hline
    \end{tabular}
\end{table*}

\subsection{BBH Properties}
In this section, we discuss the mass and orbital properties of the BBH population in more detail. 

\subsubsection{Mass and orbital separation}
One immediate result found from these simulations is that the fraction of systems that form BBHs increases for lower metallicity models; $2.4\%$, $10.8\%$, and $15.1\%$ for model $Z=0.01,\,0.001\,\mathrm{and}\,0.0001$ respectively. This is in part due to the stronger stellar winds produced in high metallicity stars which increase the amount of mass loss during the stellar evolution. This extra mass loss at Solar metallicities increases the minimum stellar mass required to form a BH and therefore the total number of BHs formed. In addition, the increased stellar mass loss leads to less massive BHs forming in general \citep{belczynski_maximum_2010}, which then receive higher natal kicks, since the fallback fraction is lower, which can more easily disrupt the binary. 
\iffalse
 with stars below this minimum mass instead, forming a NS. Since the initial mass distribution of Solar and sub-Solar models is the same, we should thus expect a greater number of NSs in the Solar metallicity model. Indeed, we do find this model to produce a total of $1.5\times$ as many neutron stars compared to the sub-Solar metallicity model. In addition, the increased stellar mass loss leads to less massive BHs forming in general \citep{belczynski_maximum_2010}, which then receive higher natal kicks, since the fallback fraction is lower, and this kick can be large enough to disrupt the binary. 
 \fi

\begin{figure*}
    \centering
    \includegraphics[width=\textwidth]{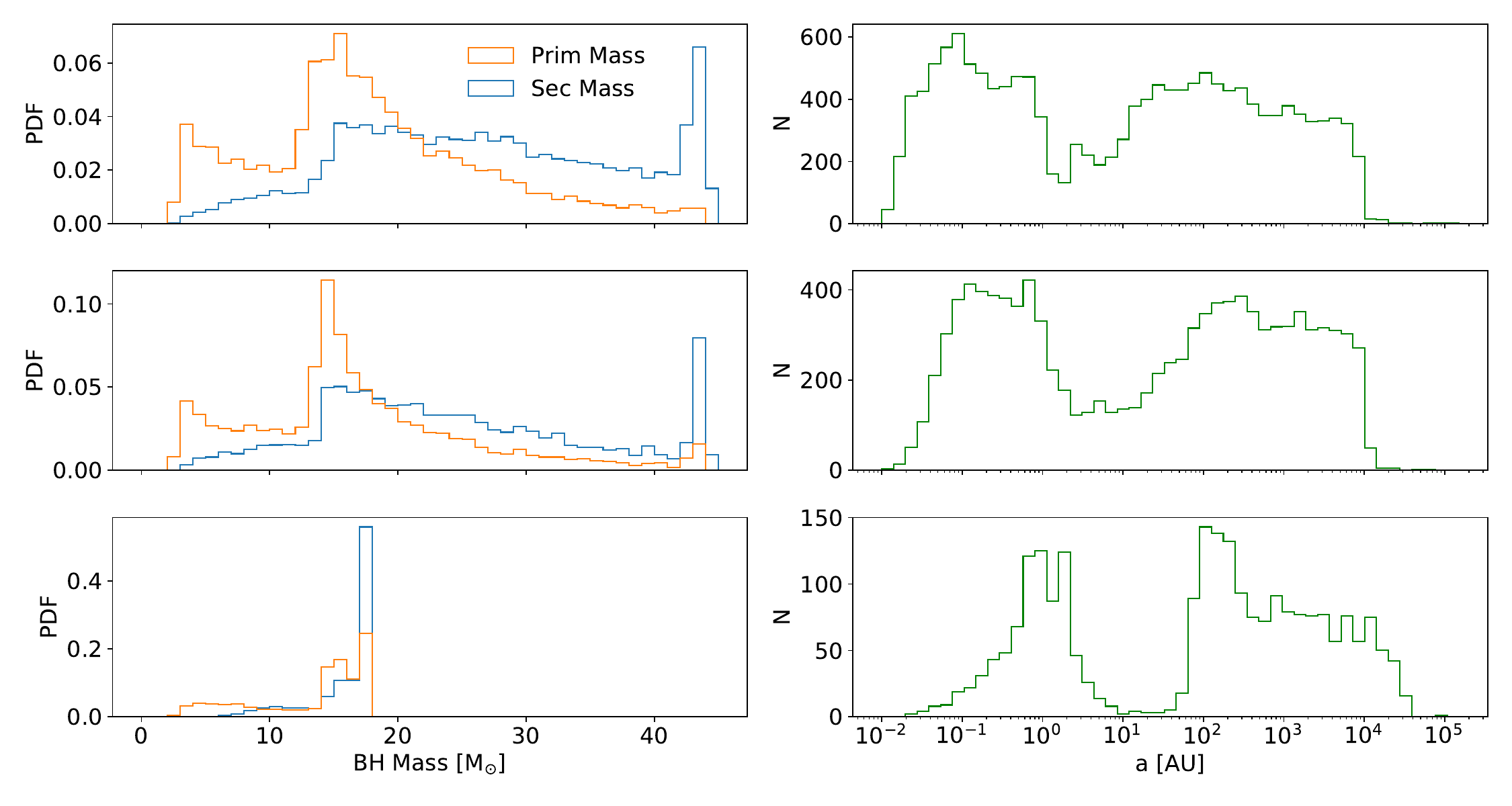}
    \caption{In the left column we show the primary (blue) and secondary (orange) BH mass distributions. The right column shows the binary separation. We plot these distributions, for three metallicity values, $Z=0.0001$ (top row), $Z=0.001$ (middle row), and $Z=0.01$ (bottom row).}
    \label{fig:massAndSemi}
\end{figure*}

The lower BH masses at higher metallicity are apparent in Fig~\ref{fig:massAndSemi}, where we see that for the solar metallicity model (bottom-left panel) $\simeq 99\%$ of all primary BHs have masses $\leq 17.3$ $M_{\sun}$. Meanwhile, both sub-solar metallicity models (top-left and center-left panels), have $50\%$ of primary BH masses exceeding $21.1 \ M_{\sun}$ and $22.9 \ M_{\sun}$, respectively. Table~\ref{tab:massdist} summarises the $50\%$, $75\%$ and $99\%$ percentiles of the primary and secondary BH mass distribution for all three models. Note that the binary components are characterised such that the primary BH is always the most massive BH.

The right column of Fig~\ref{fig:massAndSemi} shows the semi-major axis, $a$, distribution of the BBHs at formation, for each model. We see that the range of separations does not vary significantly with metallicity, always being between $\simeq 10^{-2}$ AU to $\simeq 10^{4}$ AU. However, there is a clear bi-modality in the distribution which  becomes more pronounced for higher metallicities. The bi-modality of these distributions represents the two possible pathways for forming a BBH from a primordial binary \citep{wiktorowicz_populations_2019}. BBHs at larger separations, $\gtrsim 10^{2}$ AU are formed from initially wide stellar binaries where the individual stars can evolve without much interference from each other. On the other hand, BBHs in the lower separation peak are predominantly formed following a common envelope (CE) evolution \citep{paczynski_common_1976}. This latter form of evolution occurs when one of the stars expands to such an extent that overflows its Roche lobe and begins to donate material to its companion. In the case where the receiving star is unable to accept all of the material a CE is formed, which surrounds both of the binary components. A consequence of CE evolution is the shrinking of the binary separation due to drag forces between the binary components and the surrounding envelope. Provided the separation does not shrink to the point of the stellar cores merging, this can result in the formation of the tight BBHs seen in the lower peak of the distribution \citep{livio_common_1988, xu_binding_2010, ivanova_common_2013}. It should be noted that it is possible for the stellar cores to avoid merger during the CE phase provided the envelope gets ejected before the cores have a chance to merge \citep{law-smith_successful_2022}.

\begin{table}
    \centering
    \caption{$50\%$, $75\%$ and $99\%$ percentiles for the primary and secondary BH masses for each of the three metallicity models}
    \label{tab:massdist}
    \begin{tabular}{c|c|c|c|c}
        \hline
        Metallicity & BH type & $50\%$ & $75\%$ & $99\%$\\
        & & $M_{\sun}$ & $M_{\sun}$ & $M_{\sun}$\\
        \hline
        \multirow{2}{4em}{$0.001$} & Primary & $22.9$ & $33.1$ & $44.0$\\
        & Secondary & $18.4$ & $26.5$ & $43.7$\\
        \multirow{2}{4em}{$0.01$} & Primary & $21.1$ & $30.7$ & $43.8$\\
        & Secondary & $16.1$ & $22.6$ & $43.4$\\
        \multirow{2}{4em}{$0.01$} & Primary & $17.0$ & $17.2$ & $17.3$\\
        & Secondary & $15.5$ & $17.1$ & $17.3$\\
        \hline
        
    \end{tabular}
\end{table}

\subsubsection{Eccentricity}
\begin{figure}
    \centering
    \includegraphics[width=\columnwidth]{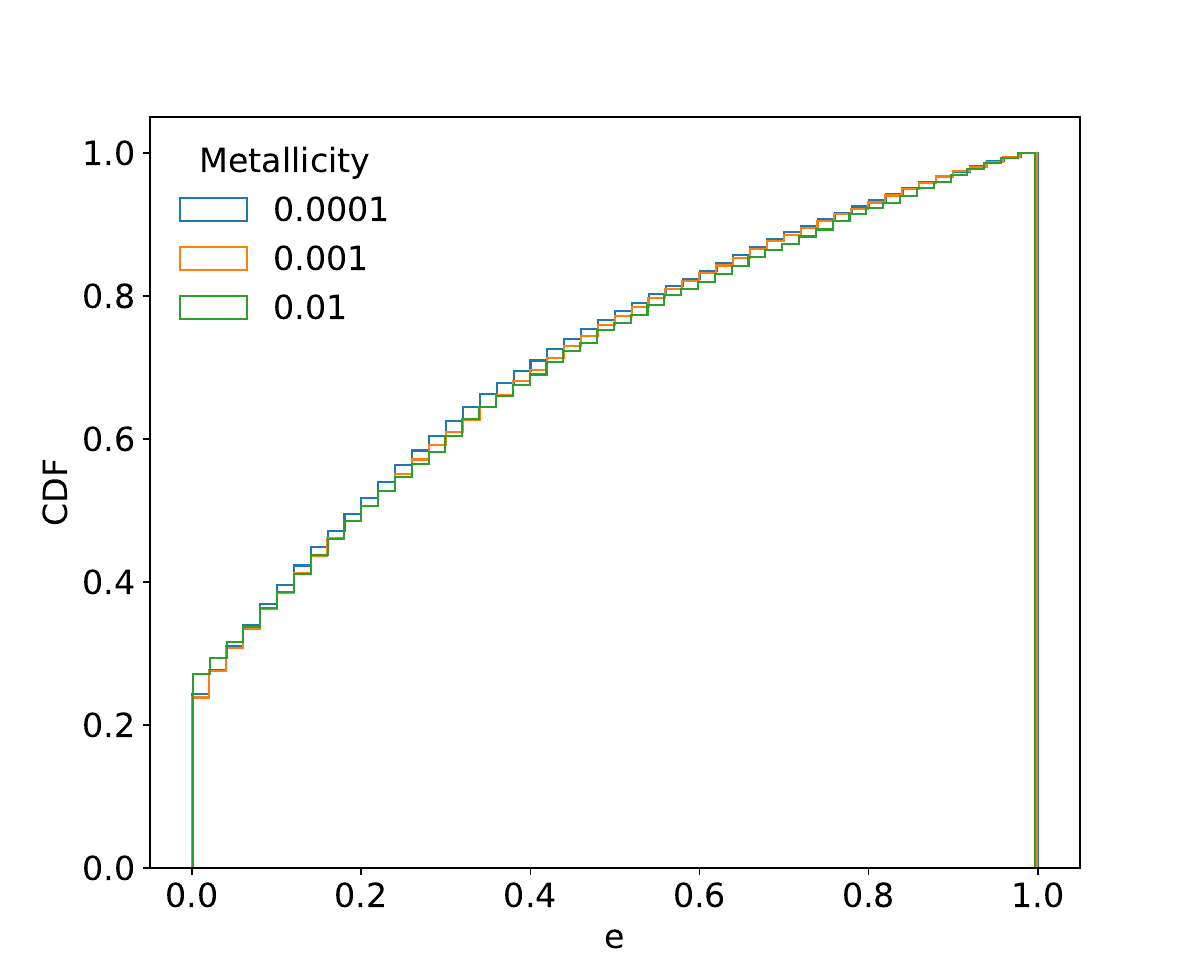}
    \caption{Cumulative distribution of the BBH eccentricities  for the three metallicity models described in the text. We can see that there is very little difference between the  models.}
    \label{fig:BBHEcc}
\end{figure}

In Fig~\ref{fig:BBHEcc} we plot the cumulative distribution of the BBH eccentricity distribution at formation. We see that the BBHs are predominantly low eccentricity with $50\%$ of BBHs having an eccentricity below $0.187$, $0.193$ and $0.161$ for metallicity models $Z=0.0001, \,0.001, \, \mathrm{and}\, 0.01$ respectively. The low eccentricities are a natural consequence of the circularising that occurs when the binary components interact through tides and mass transfer before a BBH is formed. In addition, we find that the BBH eccentricity distribution is mostly independent of the choice of metallicity. 
Although SN kicks may change the circularised binaries to become more eccentric, the fallback kick prescription ensures that most BBH progenitors have much smaller natal kick magnitudes (than their NS counterparts), preserving their smaller eccentricities.

\subsubsection{Time Delay}

\begin{figure}
    \centering
    \includegraphics[width=\columnwidth]{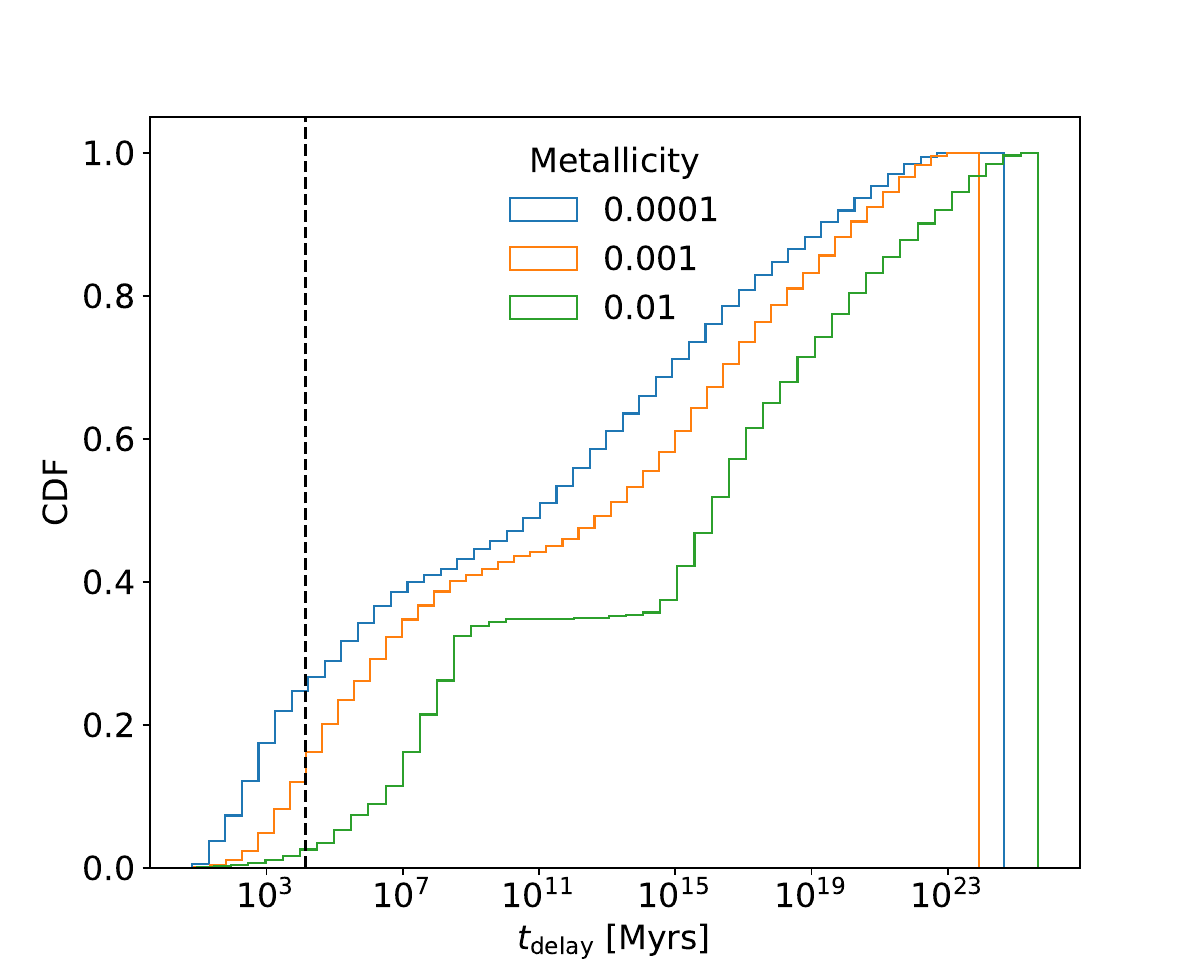}
    \caption{Time delay distribution for each metallicity model. We find that metallicity has a large impact when it comes to the number of BBH that can merge in a Hubble time. Going from low metallicity to high metallicity the fraction of BBHs with $t_{\mathrm{delay}}\leq13.7$ Gyrs is $24\%$, $12\%$ and $2.5\%$.}
    \label{fig:GWHub}
\end{figure}

%Before going any further, 
It is now helpful to look at the fraction of BBHs that would merge within a Hubble time, $t_{\rm H}$, based solely on the binary stellar evolution. To calculate this merger time or ``time delay" ($t_{\rm delay}$), we numerically integrate the merger timescale given by \cite{peters_gravitational_1964}. For a BBH of  m$_{1,2}$, eccentricity $e_0$, and semi-major axis $a_0$ in the population, 
\begin{equation}
    t_{\rm delay}(a_\mathrm{0}, e_\mathrm{0}) = \frac{12}{19}\frac{c^{4}_{0}}{\beta}\times\int^{e_{0}}_{0}de\frac{e^{29/19}[1+(121/304)e^{2}]^{1181/2299}}{(1-e^{2})^{3/2}},
    \label{eq:GWTime}
\end{equation}
where, 
\begin{equation}
    c_0 = \frac{a_0(1-e_0^2)}{e_0^{12/19}}\left [1+\frac{121e_0^2}{304}\right]^{-870/2299}
    \label{eq:c0}
\end{equation}
and,
\begin{equation}
    \beta=\frac{64}{5}\frac{G^{3}m_{1}m_{2}(m_{1}+m_{2})}{c^{5}}.
    \label{eq:beta}
\end{equation}
Fig~\ref{fig:GWHub} shows the cumulative distribution of the GW timescale for the three metallicities, along with the Hubble time $t_{\mathrm{H}}=13.7$ Gyrs line marked. We see that the higher the metallicity the lower the fraction of BBHs that can merge within a Hubble time. In particular, we find that fractionally $24\%$, $12\%$ and $2.5\%$ have $t_{\mathrm{delay}}\leq13.7$ Gyrs for metallicity models $Z=0.0001, \,0.001, \, \mathrm{and}\, 0.01$ respectively.

\section{BHs Retained in Clusters}\label{sec:retain}
\subsection{Supernovae Kicks and Escape Velocity}
To investigate the effect of stellar evolution on the BH populations inside star clusters we split the population into 3 distinct groups; single BHs, BBHs and BHs in binaries in which the other component is not a BH (i.e., it is either a star, a white dwarf or a neutron star; we denote these binaries as BH-else). We further categorise the BBH population into hard and soft binaries; where a hard binary has a larger binding energy than the average kinetic energy of the surrounding stars. Thus, by interacting with surrounding stars and COs, a hard binary will on average become harder, i.e., its binding energy  will increase. On the other hand, a soft binary will on average become softer and will eventually be dissociated by the encounters \citep{heggie_binary_1975}.

As mentioned previously, the kicks received by binary components as a result of SN are the main source of binary disruption; and even for those BBHs that remain bound a kick will still be imparted to the binary centre of mass (COM) and could be large enough to eject the binary from its host cluster.  This is an important mechanism to consider since it will set the rest of the binary's evolution. If the binary is ejected from its cluster then it would continue to evolve in the external environment (e.g., the galactic field) without further dynamical interaction.

We find that the two sub-Solar metallicity models have similar disruption fractions, with $43\%$ and $44\%$ of the initial 100,000 binaries being disrupted due to the SN kicks.
However, at solar metallicity, this fraction increases to $59\%$ of binaries being disrupted. This is due to the larger winds involved at solar metallicity which ultimately produce less massive BHs than at sub-Solar metallicities; these smaller black holes then receive higher natal kicks.

At the end of the simulation, we identify the BHs that are in still bound binaries (either with another BH or a different type of stellar object), and the BHs that are now single, after being disrupted from their initial binary. For a range of potential cluster escape velocities ($0$ to $2500$ \kms), we compare against the COM kick of binaries with a BH or the component velocity of the single BH. We are then able to estimate the fraction of retained bound BBHs and retained single BHs. %The plots in Fig~\ref{fig:retainedBHsnew} show the fraction of single BHs, BH-else, BBHs and the subset of hard BBHs that are retained for the range of escape velocities,  all normalised by the total number of BHs retained. 

In Fig~\ref{fig:retainedBHsnew} we show the fraction of BHs retained by a range of  cluster escape velocities; with the BH found as either a BBH, a single BH, or a BH-else. These are all normalised to the total number of retained BHs.% Here we omit the $0.1Z_{\sun}$ model for clarity since it was almost exactly the same as the $0.01Z_{\sun}$ model. 
We see that in the $Z=0.0001$ and $Z=0.001$ models (top and middle panels), the BBHs are the dominant form of retained BHs up to $v_{\mathrm{esc}}\sim40\,$\kms, and $v_{\mathrm{esc}}\sim20\,$\kms respectively, after which the single BHs become dominant. The BH-else binaries are a subdominant population at all values of $v_\mathrm{esc}$, with them only  contributing between $10\%$ to $15\%$ of the retained BHs. In the $Z=0.01$ model, BBHs are the dominant population only for   escape velocities $v_{\mathrm{esc}}\leq10\,$ \kms. Above this, the single BHs are the dominant population, approaching $~90\%$ of the total population at the highest escape velocities.
%At Solar metallicities, the BH-else binaries are also a subdominant source of BHs. Although, for $v_{\mathrm{esc}}>85$ \kms this population does become more significant than the fraction of BBHs. 
An important point to note is that the relationship shown in Fig~\ref{fig:retainedBHsnew} is only dependent on the escape velocity of the cluster and can therefore be applied to any cluster with that value of $v_{\mathrm{esc}}$ regardless of mass and size of the cluster.

When considering the retention fraction of the total binary (BH-BH and BH-else) and single populations we see a similar trend in the three models. Starting at the low escape velocities, the models show that the majority of the retained BHs are found in the binary population, with the single BH population becoming more dominant at higher escape velocities. This cutoff shifts slightly with the model's metallicity, with $v_{\mathrm{esc}} \gtrsim 30\,$ \kms for the $Z=0.0001$ model and $v_{\mathrm{esc}} \gtrsim 10\,$ \kms for the  $Z=0.01$ model. Since we are only dealing with stellar evolution in these models, this trend implies that in the range where the binaries are dominant, the kick velocity required to break up a typical binary is greater than the escape velocity of the cluster.

\section{Black hole binary populations}\label{sec:BBHPops}
We have discussed that for sub-Solar metallicities, the BH population for clusters with $v_{\mathrm{esc}} \lesssim 50$ \kms is predominantly in the form of BBHs, which is likely to affect the properties of the subset of merging BBHs. In addition, since the evolution of a cluster is linked to its BH subsystem \citep{breen_dynamical_2013, chattopadhyay_dynamical_2022}, the dominant presence of BBHs could impact the long-term evolution of the cluster. Thus, in what follows we investigate the properties of these {\it retained} BBHs. We subsequently look at the entire population of BBHs produced by the cluster, including ejected binaries, and consider their contribution to the merging BBH population.

\subsection{Hard binaries}
A binary is considered to be 'hard' when the binary binding energy is greater than the average kinetic energy of surrounding stars \citep{heggie_binary_1975}. From this definition, we can find an expression for a cut-off semi-major axis at the hard/soft boundary
\begin{equation}
    a_{h} = \frac{G\mu}{\sigma^{2}}
    \label{eq:hardboundary}
\end{equation}
where $\mu=m_\mathrm{1}m_\mathrm{2}/(m_\mathrm{1}+m_\mathrm{2})$ is the reduced mass of the binary and $\sigma$ is the average velocity dispersion of the cluster. The average velocity dispersion is initially proportional to the cluster escape velocity, with the exact factor dependent on the density profile assumed. We adopt the relation $v_\mathrm{esc}\approx4.77\sigma$ \citep[e.g., ][]{antonini_black_2019}, which mimics a cluster King model with $W_{0}=7$.

It is important to consider that the definition of the hard/soft boundary in Eq~\ref{eq:hardboundary} comes with some caveats. Notably, it is dependent on the distribution of energy between the binaries and singles within your cluster \citep{heggie_binary_1975} and the mass distribution of the single perturbers. Hence, we highlight that in  Eq~\ref{eq:hardboundary}, we have assumed that the BBHs have reached equipartition with a single mass population of  field stars. Another approach includes a factor of the average stellar mass of field stars $1/\langle m \rangle$, which, given the number of low-mass stars in a real cluster, would only increase the number of hard binaries (since $1/\langle m \rangle > 1$ for typical stellar mass distributions). However, this value is subject to change as the stars and whole cluster evolve. Hence, for the remainder of this paper, we will use  the definition in Eq~\ref{eq:hardboundary} and classify a binary as hard if $a<a_{\rm h}$, with the knowledge that our results assume a lower estimate for the number of hard binaries.

We now consider the sub-population of {\it hard} BBHs that is retained inside the cluster. We plot  these as a subset of the retained BBHs and as a function of the cluster escape velocity in Fig~\ref{fig:retainedBHsnew}. In all models, we see that the majority of BBHs are hard for low $v_{\mathrm{esc}}$, and that the sub-Solar model retains a high hard fraction for larger escape velocities than for Solar metallicity. In particular, we see a $>50\%$ hard fraction for $v_{\mathrm{esc}}\leq 24.1$ \kms in the $Z=0.01$ model; whereas the $Z=0.0001$ model has $>50\%$ for $v_{\mathrm{esc}} \leq 106.9$ \kms. 

The reason for the increased hard binary fraction in the sub-Solar model can be explained simply by the separation distribution shown in Fig~\ref{fig:massAndSemi}. %If we estimate the hard/soft boundary at $v_{\mathrm{esc}} = 100$ \kms, using Eq~\ref{eq:hardboundary} and assuming $m1=m2=20 \ M_{\sun}$; we find that $53\%$ of BBHs in the $0.01Z_{\sun}$ model 
% If we consider the BBH semi-major axis distributions in each of the metallicity regimes, lower panels of Fig~\ref{fig:massAndSemi}, we recall that $99\%$ of BBHs in the Solar metallicity model have a separation $a<17.3 \ \mathrm{AU}$, while almost half of the BBHs in the sub-Solar metallicity regime have wider binaries than this. Furthermore, 
We use Eq~\ref{eq:hardboundary} with $m_1=m_2=20 \ M_{\sun}$, and 
estimate a typical value for the hard/soft boundary at $v_{\mathrm{esc}}=100$ \kms; we find $a_{\mathrm{h}}=20 \ \mathrm{AU}$. For Solar metallicity,  $34.4\,\%$ of the BBHs have separations less than this and so are considered in the hard regime. On the other hand, for the sub-Solar metallicity,  $52.8\%$ of BBHs have separations below $20 \ \mathrm{AU}$. Clearly then, the sub-Solar models contain a higher proportion of tight BBHs and thus retain a large fraction of hard binaries at higher escape velocities. We also see this from the separation plot shown in Fig~\ref{fig:massAndSemi}. Although both metallicity models exhibit a bi-modality in the separation distributions, it is clear that the Solar metallicity is skewed more towards the second peak ($a>20\,\mathrm{AU}$). At $v_{\mathrm{esc}} \gtrsim 400$ \kms, while the fraction of BBH retained levels out for both metallicities in Fig~\ref{fig:retainedBHsnew},  the hard BBH fraction continues to decrease approaching zero. This is unsurprising, since the parameter $a_{\mathrm{h}}$ scales as $1/v_{\mathrm{esc}}^{2}$.
%and so, while in this high escape velocity regime none of the BBHs can escape the cluster due to their natal kicks, the velocity dispersion of the surrounding stars is continually increasing, thus the $a_{\mathrm{esc}}$, in turn, continues to decrease until none of the BBHs are tight enough to be within the hard binary regime.

The results presented so far indicate that a significant fraction of the BHs retained in a cluster will be found in binaries with another BH and that a large fraction of these binaries will have separations below the hard/soft boundary $a_{\mathrm{h}}$ (especially for clusters with $v_\mathrm{esc}<100$ \kms). Since hard binaries will on average become harder and remain bound during interactions with singles \citep{heggie_binary_1975}, we should expect that a number of them will eventually merge due to energy loss by gravitational wave radiation. Hence, they will likely contribute to the population of merging BBHs from a cluster. Moreover, they will be important to the evolution of the cluster itself as they provide an efficient energy source during the early stages of cluster evolution. 

\subsection{Binaries ejected after one dynamical encounter}\label{one-e}
The small separation of hard binaries makes the likelihood of disruption due to binary-single interactions quite low. However, their large binding energies mean that the relative recoil kick the binary receives from strong 3-body interactions can be quite large. Through consideration of energy and momentum conservation and by assuming that the average binary-single interaction increases the binding energy of the binary by some fraction $\delta$, one finds an expression of a recoil kick velocity on the binary centre of mass \citep{miller_mergers_2009}. 

\begin{equation}
    v^{2}_\mathrm{kick} \simeq \delta q_{\mathrm{3}}\frac{G}{a}\frac{m_{1}m_{2}}{m_{123}}\ ,
    \label{eq:recoilkick}
\end{equation}
where $m_\mathrm{123}=m_\mathrm{1}+m_\mathrm{2}+m_\mathrm{3}$ with $m_\mathrm{3}$ the mass of the single perturber, $q_{\mathrm{3}}=m_{\mathrm{3}}/(m_{\mathrm{1}}+m_{\mathrm{2}})$ which we assume to be $q_{\mathrm{3}}\approx0.5$. The fraction of energy given by the binary is typically averaged to $\delta=0.2$ for binary-single encounters \citep{quinlan_dynamical_1996}. 
By selecting the cases where $v_\mathrm{kick}\geq v_\mathrm{esc}$, we approximate the number of BBHs that would receive a recoil kick large enough to remove them from a cluster with escape velocity $v_\mathrm{esc}$. Within a globular cluster, we also expect binary-binary interactions, which are more complex than binary-single encounters and it is thus non-trivial to extend our recoil kick expression to binary-binary interactions.

From Eq~\ref{eq:recoilkick} we see that the harder a binary (i.e., the smaller is $a$) the larger the recoil kick experienced after an interaction.
Thus, the condition $v_\mathrm{kick}\geq v_\mathrm{esc}$ is equivalent to a condition on the semi-major axis of the binary. 
By rearranging Eq~\ref{eq:recoilkick} for $a$ and defining $a=a_{\mathrm{ej}}$ when $v_{\mathrm{kick}}=v_{\mathrm{esc}}$, the final relation for the critical semi-major axis value below which a binary is ejected is
\begin{equation}
    a_{\mathrm{ej}} = 0.1\frac{G}{v_{\mathrm{esc}}^{2}} \frac{m_{1}m_{2}}{m_{123}} \ ,
    \label{eq:aej}
\end{equation}
where $a_{\mathrm{ej}}<a_{\mathrm{h}}$ by definition.

\begin{figure}
    \centering
    \includegraphics[width=0.95\columnwidth]{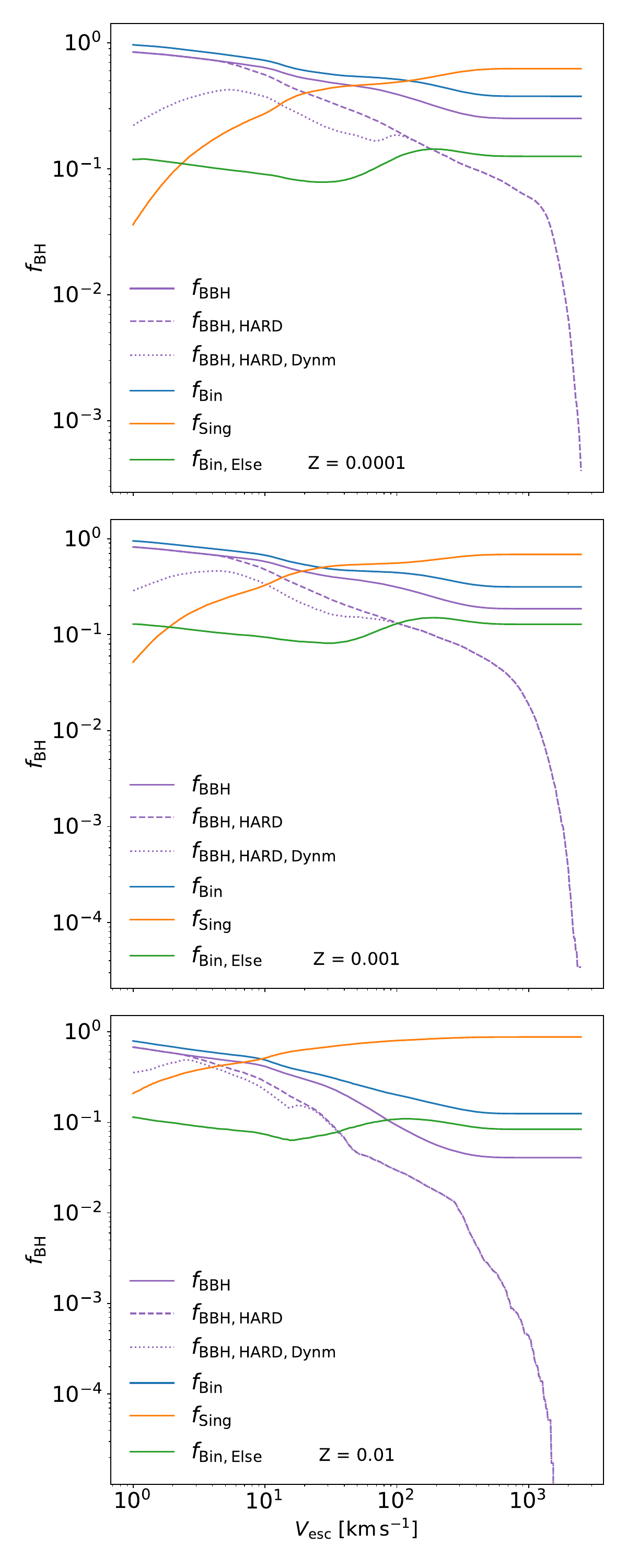}
    \caption{We show the fraction of BHs ($f_{\rm{BH}}$) retained as either single BHs (orange line, $f_{\rm{sing}}$) or as part of a binary (blue line, $f_{\rm{bin}}$). We further split the binaries into the BBH fraction ($f_{\rm{BBH}}$, solid purple line) and the binaries containing only one BH ($f_{\rm{Bin, Else}}$, green line). For the BBH group we also show the sub-fraction of \textit{hard} BBHs ($f_{\rm{BBH, HARD}}$, purple dashed line) where $a \leq a_{\rm{h}}$, with $a_{\rm{h}}$ defined in Eq~\ref{eq:hardboundary}. Finally, we show the expected number of BBHs that would remain bound to the cluster following an interaction ($f_{\rm{BBH, HARD, Dynm}}$, purple dotted line). The upper panel shows the metallicity model $0.001$, the middle panel $0.01$ and the lower panel $0.1$.}
    \label{fig:retainedBHsnew}
\end{figure}

Given that the recoil kick is also dependent on the mass of the perturber, we choose $m_{3}$ by randomly sampling the mass distribution of the retained single BHs for each escape velocity and metallicity model. In this way, we can further mimic the interactions that would likely occur if these BH populations were situated within a real cluster. The subset of hard BBHs  that are retained after a single interaction is shown in Fig~\ref{fig:retainedBHsnew} as dotted purple lines. 

Although the population of BBHs that are ejected after their first encounter, are not going to play much of a role in the overall cluster evolution, they still contribute to the population of merging BBHs. Either they were already tight enough to merge and the interaction causes the merger to occur sooner (by reducing the separation), or the decreased semi-major axis and newly drawn eccentricity caused by the interaction, now place the binary in a regime where it can merge within a Hubble time. This population would be of particular interest as their binary properties will be set mostly by stellar evolution but include some influence due to the single interaction which ejects them. The longer a primordial binary remains in the cluster, the more interactions it will experience and thus its orbital properties will become more akin to a dynamically formed binary.

When we take into account the tight BBHs that are ejected after a single interaction, we see in both metallicity models, the remaining  hard BBH population in low mass clusters goes down significantly. In the sub-Solar model, low mass clusters with $v_{\mathrm{esc}} < 7$ \kms, we see more than half of the retained hard BBHs get ejected after their first interaction. As the escape velocity increases, fewer BBHs get ejected due to their first recoil kick. This is in part because the higher escape velocity requires an equivalently large recoil kick to eject the binary; but also, as for higher escape velocities the parameter $a_{\rm h}$ gets smaller, meaning fewer BBHs are hard and so the interaction is much more likely to either widen the binary or disrupt it completely. 

In the Solar model we similarly see that the higher the escape velocity, the fewer BBHs are ejected due to the first interaction. However, we see that for $v_{\mathrm{esc}}<28$ \kms more than half of the BBHs are removed due to that first strong encounter while for $v_{\mathrm{esc}}>160$ \kms none of the first recoil kicks are able to eject the BBHs. This is slightly lower than the upper bound in the sub-Solar model where we see a few BBHs still ejected due to the interaction, up to an escape velocity $v_{\mathrm{esc}}=280$ \kms.

In all of the models,  there is a certain escape velocity above which the fraction of retained BHs that are in BBHs levels out, while the subset of those that are in hard BBHs continues to drop; this is due to the relation between the hard/soft boundary and the escape velocity of the cluster, $a_\mathrm{h}\propto 1/v^2_\mathrm{esc}$ from Eq~\ref{eq:hardboundary}. The semi-major axis cut-off for a hard binary is getting larger still and thus fewer binaries are considered hard as the escape velocity continues to increase. 

It is clear that in both metallicities models, we should expect low-mass clusters to eject a significant fraction of their hard BBH population relatively early in the cluster's evolutionary timescale. Since these ejected BBHs will typically have small separations, they are likely to make a significant contribution to the BBH merger rate for these clusters.

\subsection{In-cluster binaries unaffected by dynamics}\label{non-d}
So far we have shown that for low metallicity clusters, the BH population due to stellar evolution and simplistic dynamics is predominantly in the form of hard BBHs. Given these binaries are tightly bound, it seems probable that some fraction of systems may merge before the next strong interaction interferes with the system. To investigate this, we compare the merger timescale Eq~\ref{eq:GWTime} \citep{peters_gravitational_1964} with the interaction timescale for every hard BBH retained within various combinations of cluster mass ($M_\mathrm{cl}$) and half-mass radii ($r_\mathrm{h}$). 

\subsubsection{Timescales}
To calculate the timescale for the binaries to experience a first encounter, we again make the assumption that the encounter removes a fraction $\delta$ of the binary binding energy. We then have the rate of energy loss from the binary given by $\dot{E}_\mathrm{bin}\simeq\delta E_\mathrm{bin}/t_\mathrm{int}$ \citep{heggie_gravitational_2003}, from which we have the interaction timescale for a single binary  \citep{antonini_population_2020}
\begin{equation}
    t_\mathrm{int} \simeq \delta \frac{G m_{1} m_{2}}{2a} \dot{E}^{-1}_\mathrm{bin}
    \label{eq:interation}\ .
\end{equation}

We assume that the dynamical hardening of BBHs in the cluster core drives the cluster heating and that every binary contributes approximately the same amount of energy to the cluster. Then, we can equate the binary hardening rate to the cluster heating rate $N_{\mathrm{bin}}\dot{E}_\mathrm{bin}=\dot{E}$, where $N_{\mathrm{bin}}$ is the number of binaries in the cluster. From which we can further relate to the heat generation to the global cluster properties  \citep{henon_sur_1961,  breen_dynamical_2013}
\begin{equation}
    \dot{E}=\zeta \frac{|E|}{t_\mathrm{rh}},
    \label{eq:energyGeneration}
\end{equation}
where $E\simeq-0.2GM^{2}_\mathrm{cl}/r_\mathrm{h}$ is the total energy of the cluster and  $\zeta\approx0.2$ \citep{henon_sur_1961, henon_exploration_1965, gieles_life_2011}. The cluster relaxation time is given by
\begin{equation}
    t_\mathrm{rh} = 0.138\sqrt{\frac{M_\mathrm{cl}r^{3}_\mathrm{h}}{G}}\frac{1}{\langle m_\mathrm{all} \rangle \psi \ln \Lambda}\ ,
    \label{eq:relax}
\end{equation}
where $\langle m_{\rm all} \rangle=0.809$ M$_{\odot}$ is set to the average stellar mass initially in the cluster, which is calculated using a \citet{kroupa_variation_2001} IMF between $0.08$ and $150$ M$_{\odot}$. $\ln \Lambda$ is  the Coulomb logarithm which we set to a constant $\ln \Lambda = 10$, and $\psi$ depends on the mass spectrum within the half-mass radius. For a single component cluster, $\psi=1$, but in what follows, we adopt $\psi=5$. This takes into account that in the early evolution of the cluster (first $\sim 100$Myr) the mass function contains more massive stars and $\psi$ is high.

Combining Eq~\ref{eq:energyGeneration} and Eq~\ref{eq:interation} we arrive at an expression for the expected total timescale between all interactions, $t_\mathrm{int}$, in terms of the cluster half-mass relaxation time ($t_\mathrm{rh}$) \\

\begin{equation}
    {t_\mathrm{int}}\simeq 25\delta\frac{m_{1}m_{2}}{M_\mathrm{cl}^{2}}\frac{r_\mathrm{h}}{a} N_{\mathrm{bin}} {t_\mathrm{rh}}\ .
    \label{eq:interactionTime}
\end{equation}
We note that the derivation of Eq~\ref{eq:interactionTime} makes the assumption that the cluster has undergone several relaxation times, such that it has reached a state of balanced evolution \citep{henon_sur_1961, breen_dynamical_2013}. Once in this state, we can relate the heat generation due to BBHs in the cluster core to the global properties of the cluster itself as in \citet{breen_dynamical_2013}. 
As before, we set $\delta=0.2$ which is the expected averaged value for binary-single interactions. However, we note that  $\delta$ should be a distribution of values, and that the average is expected to be somewhat higher for binary-binary interactions \citep{zevin_eccentric_2019}. Here, we ignore these complications and continue with a fixed value.

It is important to consider that our assumption of all primordial binaries contributing equally to the heating of the cluster all of the time is likely not realistic. It is more feasible that only a fraction of the primordial binaries are directly contributing to the heating at any given time, and so our assumption is producing a conservative, lower estimate for the interaction timescale. As a comparison we completed the analysis shown in the following section also assuming that none of the primordial binaries reach the core and so the interaction rate can simply be computed as \citep[e.g.,][]{spitzer_dynamical_1987}

\begin{equation}
    \begin{split}
        t_{\mathrm{int}} \approx 2\times10^{7} \zeta^{-1} \left(\frac{n}{10^{6}\,\mathrm{pc^{-3}}}\right)^{-1}\left(\frac{\sigma}{30\,\mathrm{km\,s^{-1}}}\right)   \left(\frac{m_{\mathrm{*}}}{m_{\mathrm{3}}}10\right)^{1/2} \\ 
        \left(\frac{a}{0.04\,\mathrm{AU}}\right)^{-1}\left(\frac{m_{\mathrm{12}}}{20\,\mathrm{M_{\sun}}}\right)^{-1}\,\mathrm{yr} .
    \end{split}
    \label{eq:altTint}
\end{equation}
Here we have the number density of the cluster, $n$, the average star mass in the cluster, $m_{\mathrm{*}}$, and the constant $\zeta \leq 1$ which parameterizes the difference from cluster equipartition (we set $\zeta=1$ for this simple check). Using this alternative approach to the interaction timescale we found very little effect on the population fractions of the BBHs defined in the following section (Fig~\ref{fig:popfractions}). Similarly, we only found only slight differences in the merging population (Fig~\ref{fig:mergefractions}). Since we would draw the same conclusions on the BBH populations using either approach for interaction timescale, we continue this work assuming $t_{\mathrm{int}}$ as defined in Eq~\ref{eq:interactionTime}.
%The resultant gravitational wave timescale is then scaled by the relaxation time for the various clusters, where we take the half-mass relaxation time as defined by \citet{spitzer_random_1971}.

\subsection{Merging population}
With the consideration  made in the previous sections, we are able to make the distinction between three distinct BBH populations. The first population, Pop I, are the tightest binaries, and they experience no dynamical encounters before they merge. These BBHs fall into one of two categories. One possibility is that they are ejected from the cluster by the SN kick of one of the binary components, continuing to evolve in the galactic field until they merge within a Hubble time. The other option is that the BBH remains in the cluster, however, its GW timescale is shorter than the typical interaction timescale of the cluster, hence it will merge before it can experience a strong encounter that significantly affects the binary properties (Section \ref{non-d}). Clearly, this population of merging BBHs should closely resemble that of the isolated binary formation channel, since its properties are solely dictated by the stellar evolution of the binary.

We define Pop II BBHs as those that will experience a single strong interaction\footnote{We note that for these COMPAS models we do not consider the effect of 3-body interactions prior to the BH formation.} that ejects them from the cluster, i.e., $a \leq a_{\mathrm{ej}}$ (Section \ref{one-e}). Meanwhile, Pop III are the remaining \textit{hard} BBHs that will experience multiple strong encounters inside the cluster, i,e., $a_{\mathrm{ej}} < a \leq  a_{\mathrm{h}}$. % Depending on the number of encounters that the Pop III BBHs experience before they merge or are ejected, their binary properties will become more akin to the dynamically formed population. However, since Pop II is defined as having only a single interaction before they are ejected; they will always be a hybrid BBH population where the binary properties are mostly set by stellar evolution, but the single strong encounter is likely to leave some imprint on the binary.

\begin{figure*}
    \centering
    \includegraphics[width=\textwidth]{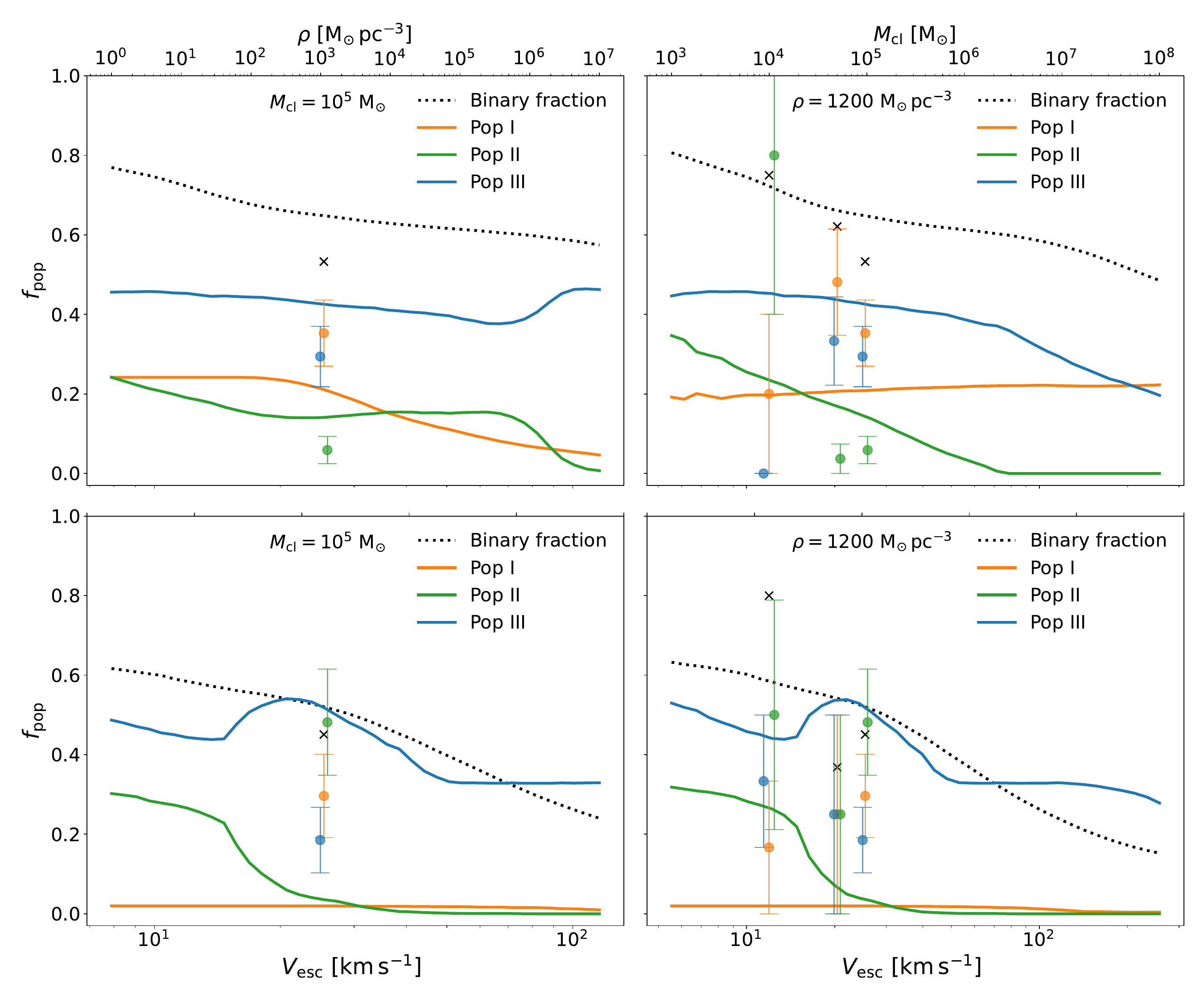}
    \caption{Here we show the primordial BBH population split into three sub-populations based on the binary separation. Pop I (orange lines) are BBHs that experience no interactions before they merge (either outside the cluster or before an encounter in the cluster). Pop II (green lines) are BBHs that experience one strong interaction that ejects them from the cluster and Pop III (blue lines) are hard BBHs that experience more than one encounter in the cluster. In addition we show the overall retained BH-binary (binary with at least one BH) fraction (black dashed line) across the range of escape velocities. In the left panels we fix the cluster mass at $M_{\mathrm{cl}} = 10^{5} \ \mathrm{M_{\sun}}$ and vary the cluster density from $\rho = 1 \ \mathrm{M_{\sun}\,pc^{-3}}$ to $10^{7} \ \mathrm{M_{\sun}\,pc^{-3}}$. The right panels show a fixed density of $\rho = 1200 \ \mathrm{M_{\sun}\,pc^{-3}}$ while varying the cluster mass from $M_{\mathrm{cl}} = 10^{3} \ \mathrm{M_{\sun}}$ to $10^{8} \ \mathrm{M_{\sun}}$. We show the results for a sub-Solar metallicity model ($Z=0.0001$) in the top panels, and for a Solar metallicity model ($Z=0.01$) in the bottom panels. Finally, the coloured points represent the corresponding populations as found in the $N$-body models and similarly the black crosses are the retained BH-binary fraction of the cluster.}
    \label{fig:popfractions}
\end{figure*}

Fig~\ref{fig:popfractions}, shows the fraction of BBHs split into these populations across a range of cluster escape velocities. %$4.42$ \kms $ \ \rightarrow 1.43 \times 10^{3}$ \kms. 
%These plots can also be interpreted as an increasing cluster density from order $10^{2} \ M_{\sun} \mathrm{pc}^{-3}$ up to order $10^{5} \ M_{\sun} \mathrm{pc}^{-3}$.
%The upper panel shows the sub-Solar metallicity model, where we see that for $v_{\mathrm{esc}} > 6.56$ \kms the Pop III BBHs are the dominant population, whilst the Pop I and Pop II BBHs account for a similar fraction of binaries for $v_{\mathrm{esc}} \gtrsim 20$ \kms. 
It should be noted that one extra population of BBHs that is not plotted here are the soft BBHs, where $a>a_{\rm h}$, since these binaries are likely to be disrupted and contribute to the single BH population. However, we know from Fig~\ref{fig:retainedBHsnew} that these become the dominant form of BBHs in very massive clusters with high escape velocity. The definition of these populations relies on the calculation of the cluster interaction timescale, which is dependent on both the cluster mass and the half mass density. Therefore we are unable to simply plot against the cluster escape velocity as we had previously done in Fig~\ref{fig:retainedBHsnew}. Instead, we show relationship between the fraction of these populations against varying cluster half mass density with fixed cluster mass, $M_{\mathrm{cl}}=10^{5}\,\mathrm{M_{\sun}}$ (left panels) and vice-versa with fixed density of $\rho=1200\,\mathrm{M_{\sun}pc^{-3}}$ (right hand panels). For a given cluster mass and density, we also compute the cluster escape velocity which is plotted on the main x-axis (with the corresponding varying cluster mass and half mass density shown on the secondary x-axis).

%The lower panel of Fig~\ref{fig:popfractions} shows the BBH population fractions for the Solar metallicity model. In this case, the Pop II BBHs are the dominant population for $v_{\mathrm{esc}}<34.3$ \kms. Once this escape velocity is reached, Pop II accounts for $43.5\%$ of the BBHs, while Pop III account for $44.0\%$ and Pop I accounts for $11.7\%$. Above this escape velocity, the contribution of Pop I and Pop II BBHs rapidly decreases, such that by $v_{\mathrm{esc}}=135$ \kms, Pop I accounts for $3.05\%$ BBHs and we find no Pop II binaries. At this escape velocity, we also see a maximum in the contribution from Pop III, where they account for $93.3\%$ of the binaries. As we move to even higher escape velocities, the Pop III binaries also decrease in number, since the $a_{\rm h}$ cut-off decreases correspondingly and the soft binaries become the dominant BBH population.

\begin{figure*}
    \centering
    \includegraphics[width=\textwidth]{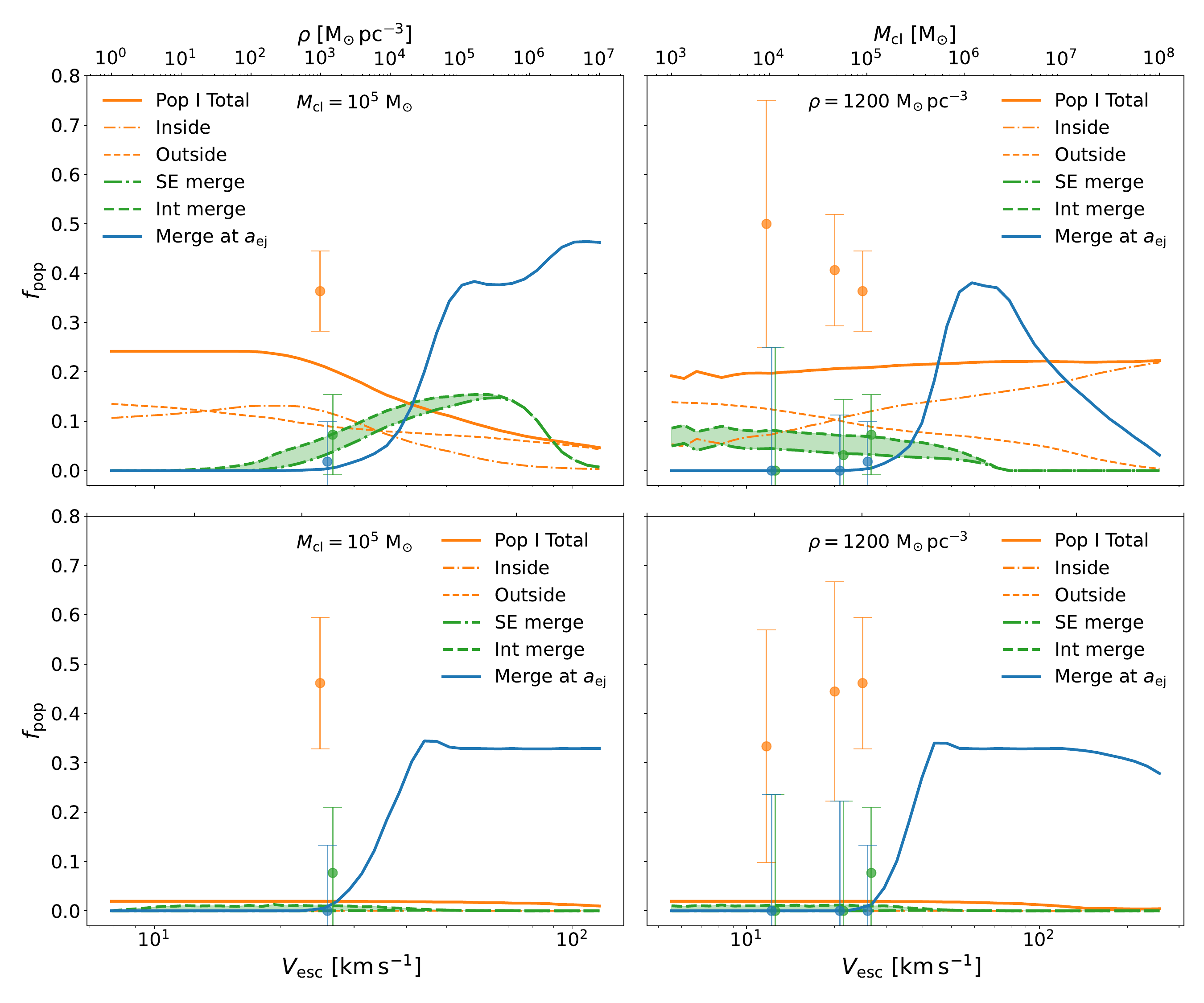}
    \caption{We show the fraction of merging BBHs normalised by the total number of BBHs, split between three populations (Pop I, Pop II and Pop III) defined in the text and caption of Fig~\ref{fig:popfractions}. For Pop I mergers (orange lines), we show the number of mergers inside the cluster (dashed-dot lines), outside the cluster (dashed lines) and the total number (solid lines). Since Pop II experience a single encounter which ejects them from the cluster, we compute two limits for the mergers. The first a lower estimate assuming the orbital properties from the stellar evolution (SE merge, dashed-dot line), and the then an upper limit assuming a single interaction which increases the binding energy by $20\%$ and produces an eccentricity kick which is drawn from a thermal distribution (Int merge, dashed line). Between these limits we show the green shaded region. Finally, for Pop III mergers we compute an upper estimate, assuming that the BBHs undergo multiple interactions until the separation shrinks to $a_{\mathrm{ej}}$. At which point we compute the effect of the interaction on the binary orbital properties. We show (in corresponding colours) the results from our $N$-body simulations, including both the mergers that occur within the simulation time (1 Gyr), and those escaped systems that would merge within a Hubble time according to Eq~\ref{eq:GWTime}.  
    }
    \label{fig:mergefractions}
\end{figure*}

It is possible for all three of these populations to contribute to the sub-population of merging BBHs under slightly different conditions, and thus we investigate how the contribution of each population changes across the range of escape velocities used in Fig~\ref{fig:popfractions}. By definition, Pop I BBHs all merge either before an interaction or outside of the cluster after being ejected by the SN kicks, hence in Fig~\ref{fig:mergefractions} we divide up the Pop I BBHs into outside and inside mergers. For a sub-Solar metallicity model, $Z=0.0001$ and fixed $\rho$ (upper right panel), we see that at low escape velocities, $v_{\mathrm{esc}}<21.3$ \kms, the Pop I mergers are dominated by outside mergers. Meanwhile, in the Solar model with fixed $\rho$ (lower right panel), the outside Pop I mergers remain dominant up to $v_{\mathrm{esc}}=191.5$ \kms, above which point the mergers from Pop I become dominated by inside mergers. This difference between metallicities is likely a result of the larger SN kicks imparted on the BBHs at Solar metallicities. This means that it is much easier for the Pop I binaries to be ejected at lower escape velocities in the Solar metallicity case, thus leading to more outside mergers. When the cluster mass is kept constant (left panels) we see that in the Solar model (lower panels) the Pop I mergers are always outside of the cluster. On the other hand, for the sub-Solar model (upper left panels) the outside mergers are almost always dominant.

For the other two populations, Pop II and Pop III, we must calculate what fraction of them will undergo a merger within a Hubble time and thus how they will contribute to the merging population. Recall that we define Pop II BBHs as binaries whose semi-major axis is smaller than some cut-off value, $a<a_{\mathrm{ej}}$, where $a_{\mathrm{ej}}$ is defined by Eq~\ref{eq:aej} and describes the separation at which a single strong interaction ejects the BBH from the cluster. Naturally then, for this population to merge either the binary properties as set by the stellar evolution place the BBH in a merging regime or the single strong interaction adjusts the properties such that the BBH can now merge in a Hubble time. Assuming either of these scenarios we set upper and lower limits on the number of expected mergers from Pop II.

We first estimate the lower limit of mergers from Pop II by assuming the interaction does not affect the binary properties and so they are set solely by the SE. With these parameters we calculate the GW timescale from Eq~\ref{eq:GWTime}, and check how many would merge within a Hubble time. For the upper limit, we account for the effect the single strong interaction has on the binary. We assume that the encounter reduces the binary binding energy by $20\%$ and draw a new eccentricity by averaging 10 random samples from a thermal eccentricity distribution. The upper limit for mergers can then be found by how many merge within a Hubble time. We show these limits in Fig~\ref{fig:mergefractions} (the green lines). Note that the values shown in Fig~\ref{fig:mergefractions} are normalised by the total number of BBHs across the simulation since the total fraction of merging BBHs is dependent on whether you take the upper or lower limit for the Pop II mergers. In the sub-Solar model, when the cluster mass is kept constant, we see that the fraction of Pop II mergers gradually increases until a maximum of $18\%$ at $76$ \kms after which is rapidly drops towards zero. This increase in dominance is likely due to the shortened interaction timescale at higher densities. For some of the tight binaries that remain in the cluster, this means that they are no longer able to merge before an interaction timescale. Essentially this is a shift from "inside" Pop I mergers to Pop II mergers.

This is supported by the fact that we see the those "inside" Pop I mergers drop to zero at the same time as Pop II grows. However, this growing number of Pop II mergers will always be turned around if the density (and thus the escape velocity) continue to increase, since $a_{\mathrm{ej}}\propto\frac{1}{v_{\mathrm{esc}}^{2}}$. Therefore, since the $a_{\mathrm{ej}}$ cut-off value decreases for larger $v_{\mathrm{esc}}$ the Pop II binaries close to $a_{\mathrm{ej}}$ will fall into the Pop III group as they no longer receive a large enough kick to escape the cluster on a single interaction. When we instead keep the density constant (upper right panel) we see that the fraction of Pop II mergers has a fairly consistent slight downward trend up to $100\,$\kms after which point it drops quickly to zero. This difference with the fixed mass case discussed previously can likely be explained by the different trends we see in Fig~\ref{fig:popfractions}. There we see that the fixed density case sees the Pop II fraction simply start at $~40\%$ of the BBH population and then quickly fall to zero as the cluster mass increased. On the other hand the fixed mass case (upper left) sees the Pop II fraction hover around $~20\%$ of the BBH population across almost the entire range of densities considered. Interestingly, this difference in trend is not particularly seen in the Solar model (lower panels of Fig~\ref{fig:popfractions}) where instead fixing either the cluster mass or the density has negligible effect on the trend of the Pop II fraction. This then is carried through to the mergers in the Pop II group, where we see at Solar metallicity (lower panels of Fig~\ref{fig:mergefractions}) the same distribution of Pop II mergers when we fix density and mass respectively.

In the case of Pop III mergers, they exist in the cluster for more than a single interaction and since they are still hard binaries we can assume that each successive interaction shrinks the semi-major axis, until it reaches $a_{\mathrm{ej}}$ at which point the subsequent interaction ejects the binary. In the process of the many encounters leading to $a_{\mathrm{ej}}$ it is possible that the that one of the interactions leads to a merger before the binary reaches $a_{\mathrm{ej}}$. However, it is difficult to consider this as it is very dependant on each individual encounter. Thus we opt to compute a lower limit on the mergers. Assuming that all of the Pop III binaries are able to shrink to $a_{\mathrm{ej}}$ without merging earlier, we then consider the final encounter in the same way as for Pop II mergers, and see how many mergers occur within a Hubble time. It should be noted that we do account for the number of interactions that it will take to reach $a_{\mathrm{ej}}$ from the binaries initial separation, and this is factored into the time to merger. Although this population originates from the primordial population, the longer they remain within the cluster the more interactions they experience which will change the binary orbital parameters. Over enough encounters the orbital properties of the binary would more closely resemble that of a purely dynamically formed BBH. However this theoretical treatment of potential interactions has a couple of caveats. Firstly, we do not consider 3-body interactions on a binary until it has formed a BBH. Secondly, we do not consider exchanges of binary components from these encounters, which would in turn mean a change in the component mass.

In Fig~\ref{fig:mergefractions} we also show the fraction of mergers coming from Pop III, and we see that in both models (and in both variations of fixed $\rho$ and fixed $\mathrm{M_{cl}}$) this population doesn't contribute to the mergers until $\approx25\,$\kms. In the Solar models (lower panels), Pop III quickly becomes the dominant contributor to the mergers (by $\approx40\,$\kms). In addition, we see that at Solar metallicity, both when fixing density and cluster mass, the contribution from Pop III eventually levels out at $\approx 32\%$ of the initial BBHs formed. At  sub-Solar metallicity (upper panels), we see that while Pop III does eventually become the dominant source of mergers there is still a significant amount of Pop I and Pop II mergers.

\subsection{Comparison to $N$-body simulations}

So far we have assumed a simplistic cluster model where we only start to consider the dynamics of the cluster after the stars have evolved and formed the BH populations. This is useful to estimate the effect of the stellar evolution on the BBH population and specifically on the sub-population of merging BBHs. However, in reality, the dynamics of the cluster during the period of stellar evolution may impact the  BH populations. To investigate this we utilise the high-performance hybrid N-Body code, PeTar \citep{wang_petar_2020, nitadori_accelerating_2012}, which allows us to populate a star cluster with some given density profile, and evolve the stars (both single and those in binaries) whilst still considering the dynamical interactions of the surrounding cluster. In comparison to direct N-body codes, PeTar combines the particle-tree particle-particle method \citep{oshino_particleparticle_2011} and the slow-down algorithmic regularisation method (SDAR) \citep{wang_sdar_2020} with parallelisation using a hybrid parallel method based on the FDPS framework \citep{iwasawa_implementation_2016,iwasawa_accelerated_2020, namekata_fortran_2018}. This allows the simulations to be much quicker than other direct N-body codes whilst also giving us the option to simulate massive star clusters with binary fractions approaching 100\% \citep{wang_impact_2021}. Stellar evolution in PeTar follows the updated single and binary stellar evolution packages \citep{banerjee_bse_2020, hurley_comprehensive_2000} where we choose all of the stellar parameters to mimic those used in the COMPAS runs. 

We run 6 different cluster models, 3 at the sub-Solar metallicity $Z=0.0001$ and 3 at Solar metallicity, for three different cluster masses; $M=10^{4} \ M_{\sun}$, $5\times10^{4} \ M_{\sun}$ and $1\times10^{5} \ M_{\sun}$. For all these models we keep the density fixed at $\rho_{\rm h} \approx 1200 \ M_{\sun}\mathrm{pc}^{-3}$. Each cluster is initialised with a \citet{king_structure_1966} model where the concentration parameter is set to $W_{0}=7$ and the stellar masses drawn from a \citet{kroupa_variation_2001} IMF with a range of $0.08 \ M_{\sun}$ to $150 \ M_{\sun}$. We then set the initial binary fraction such that all stars $M>20 \ M_{\sun}$ are placed in a binary, with the partner star randomly selected from a uniform q-distribution between 0.1 and 1. We then adjust the binary period and eccentricity according to the extended \citet{sana_binary_2012} distribution described in \citet{oh_dependency_2015} which matches the adjustment made for the binaries in the COMPAS models.

Each model is simulated for 1 Gyr which gives more than enough time for all of the primordial binaries to either form a compact object binary or to be disrupted due to the interactions, and for the clusters to have significantly evolved through dynamics. From the data, we extract the binary information at the time that each BBH is formed and run the same population tests as we did for the COMPAS models  to separate these BBHs into the same population groups. The key difference here is that by using a self-consistent cluster model we can take into account the dynamical interactions within the cluster, as well as the evolution of the cluster itself before the formation of the BHs.
%we can use the actual properties of the cluster; mass, half-mass radii, and an escape velocity gradient, at the time each binary forms a BBH 
This allows us to get a more accurate picture of  these populations in a more realistic setting. The cluster mass and half-mass radii can simply be read from the data at the time of the BBH formation; however, to calculate the escape velocity for a specific binary at this time, we must take into account its position within the cluster and find the average velocity dispersion of the surrounding stars. From this we can find the escape velocity of the cluster using the relation $v_{\mathrm{esc}}\simeq4.77\sigma$ for a King density profile with $W_{0}=7$ (as used for the COMPAS Model). 

We show the results of these N-body simulations as markers on Fig~\ref{fig:popfractions} and Fig~\ref{fig:mergefractions}. These markers are not in a one-to-one correspondence with the COMPAS results, since the binary stellar evolution code used is subtly different in PeTar compared to COMPAS. It is also important to note that the uncertainty in the data for the smallest cluster we simulated, at $v_{\mathrm{esc}}=14$ \kms, is very large since we are dealing with a relatively low number of statistics here, less than ten binaries. From Fig~\ref{fig:popfractions}, one of the key differences we see between the PeTar results and the COMPAS results is the slight reduction in the number of pop III binaries. This suggests some dynamical interactions during the stellar evolution phase of the binaries, disrupt some of these wider BBHs. 

\begin{figure}
    \centering
    \begin{subfigure}{\columnwidth}
    \centering
        \includegraphics[width=\columnwidth]{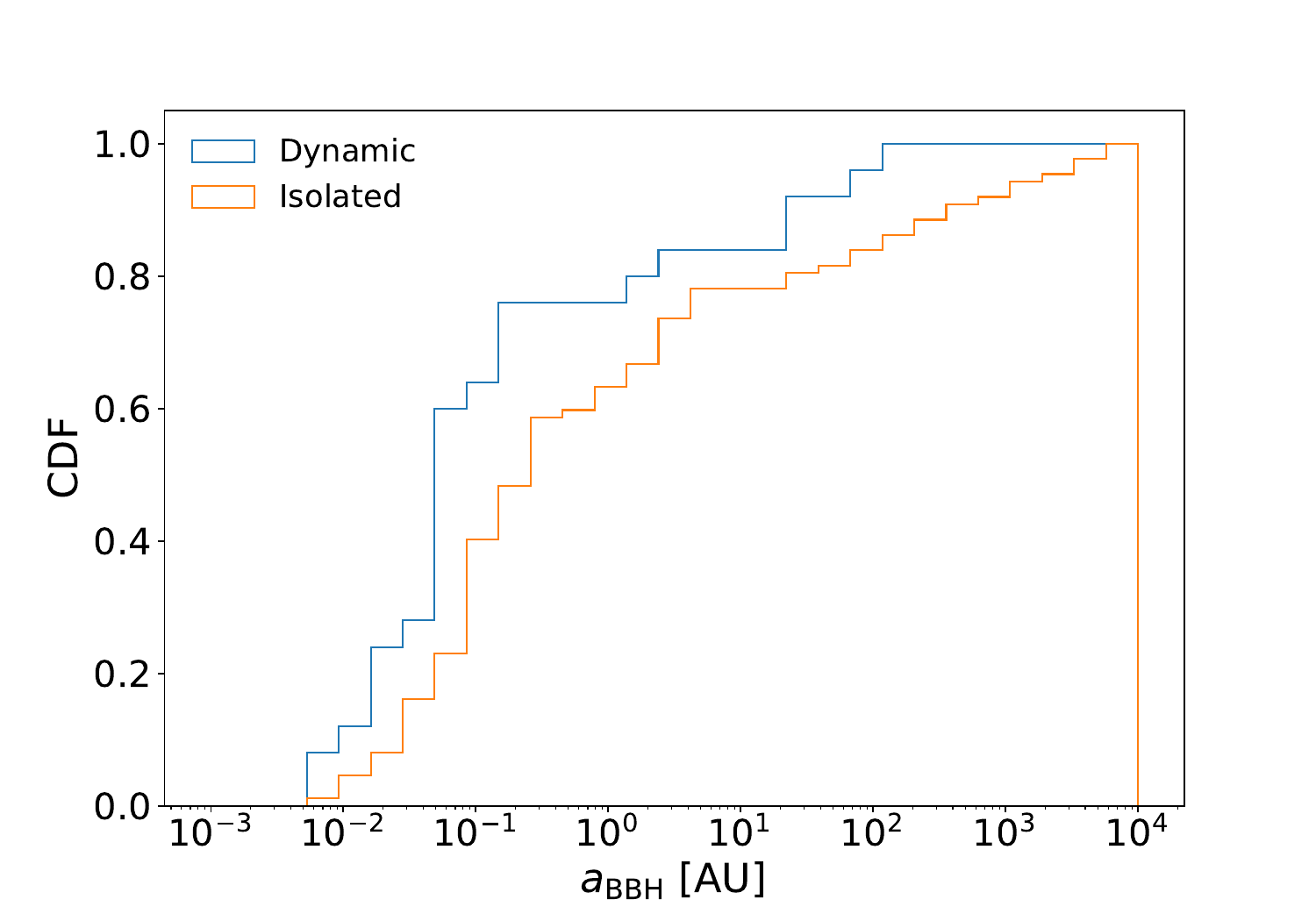}
    \end{subfigure}
    
    \hfill
    
    \begin{subfigure}{\columnwidth}
    \centering
        \includegraphics[width=\columnwidth]{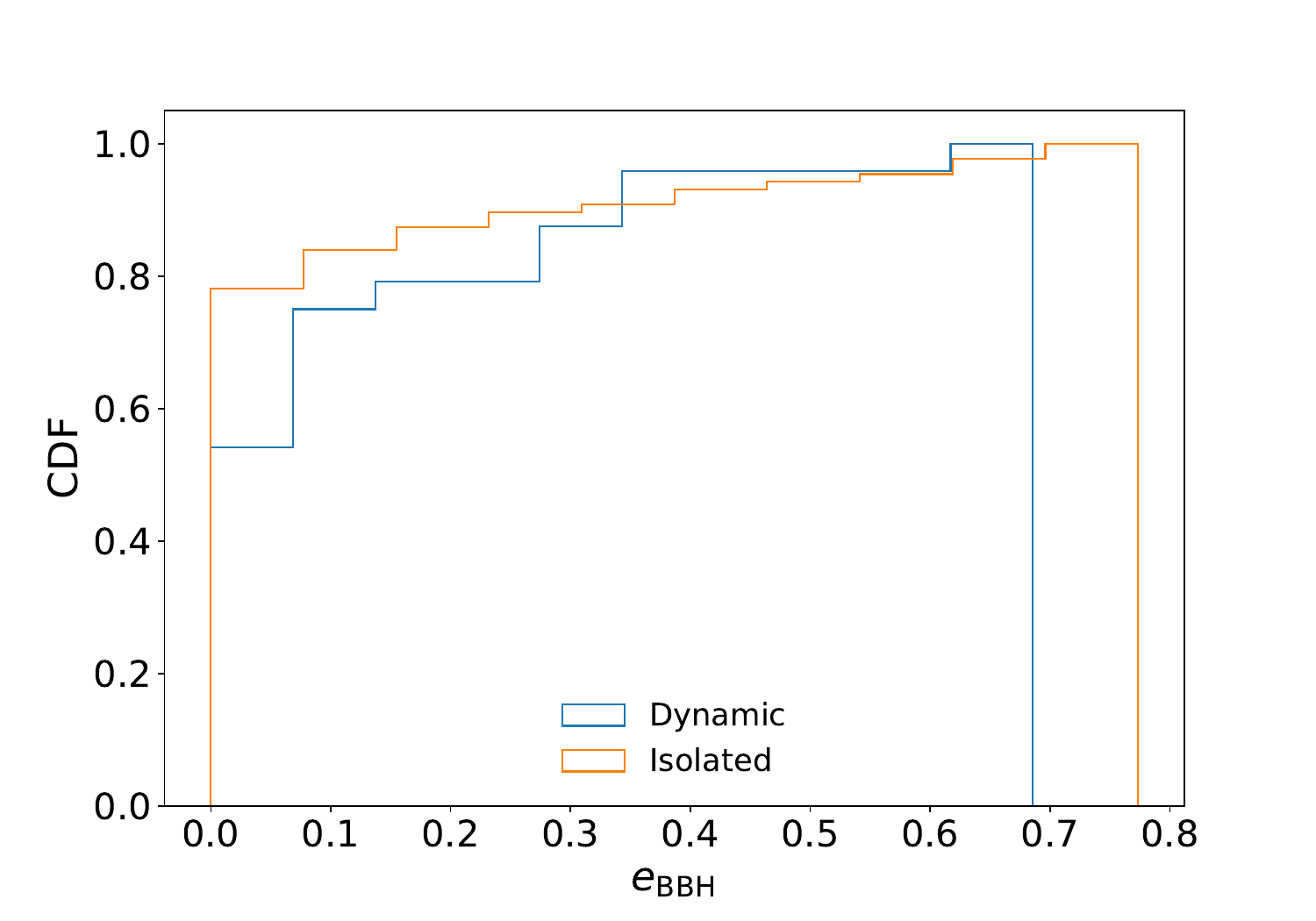}
    \end{subfigure}
    
    \caption{We compare the separation (upper panel) and eccentricity (lower panel) distributions when considering only binary stellar evolution, and in the context of a dynamical environment. We see that there is very little effect on the eccentricity of the formed BBHs; however, the separation of the BBHs is typically reduced when dynamics are introduced. These plots show results only from the Solar metallicity models.}
    \label{fig:NbodyVSisol}
\end{figure}

In order to better quantify the effect of dynamics on the binary population during the period of stellar evolution, we use the independent binary stellar evolution tool in PeTar to evolve a population of binaries drawn from the same initial distributions as those used in the cluster simulations. We then compare the separation and eccentricity distributions for the formed BBH in each case (with dynamics and isolated); these comparisons are shown in Fig~\ref{fig:NbodyVSisol}. From the comparison of the separation distributions it is clear that the dynamical environment causes two main effects; most importantly, the disruption of the wider binaries $\gtrsim 10\, \mathrm{AU}$ and secondly, the hardening of the tighter binaries. This results in the separation distribution shifting to smaller $a_{\mathrm{BBH}}$. The eccentricity distribution is not overly affected by the introduction of dynamics.

These N-body runs have shown us that, dynamical encounters prior to the formation of BHs tend to slightly reduce the prevalence of the wider Pop III binaries. However, these still account for at least $\sim 20\%$ of the BBH population In our models (excluding the single case of 0 binaries in the sub-Solar metallicity model in which we are dealing with very low number statistics $<5$). Therefore, it still seems that our method for splitting the BBH population is applicable even when considering early dynamics in the cluster.

In addition, we investigate the number of actual mergers that occur in the simulations until the final integration time of $1\rm Gyr$, and including  
those that occur within the ejected population in less than the Hubble time. 
%These simulations utilise a slight variation on the stellar evolution parameters, namely including Bondi-Hoyle wind accretion \citep{comerford_bondi-hoyle-lyttleton_2019}, however this should not strongly affect the mergers. 
After identifying  the mergers in the simulations, we then look back to the time of BBH formation for each of these binaries so that we can characterise them into the three populations. In addition, for BBHs that still exist at the end of the simulation and have escaped the cluster, we calculate the time delay using Eq~\ref{eq:GWTime}, and check if they would merge within a Hubble time. We plot these points on top of the COMPAS results in Fig~\ref{fig:mergefractions} normalised to the initial number of BBHs that are formed from the primordial population; and also quote the merger counts in Table~\ref{tab:N-bodyMergers}. As was the case for Fig~\ref{fig:popfractions}, the markers on Fig~\ref{fig:mergefractions} are not in a one-to-one correspondence with the COMPAS.
%and they are shown here more for as a reference.

We see a clear dominance of mergers from Pop I in every model, especially for the most massive $10^{5} \ M_{\sun}$ cluster where the errors are smaller. We also note that we find a significantly higher fraction of Pop I mergers in the N-body runs than we had predicted from the COMPAS models. Pop III mergers are non-existent within the solar metallicity models, and there is only a single merger found in the most massive cluster at sub-solar metallicities. Pop II mergers are also very sub-dominant, always around 5 times fewer than Pop I mergers.
%This is somewhat surprising since from Fig~\ref{fig:popfractions} we see that for Solar metallicity clusters the fraction of Pop I and Pop II BBHs in the cluster are similar.Although, Pop II binaries are not defined to merge, like that of Pop I, one still might have assumed a smaller gap between the merging sub-populations than what we see in Fig~\ref{fig:mergefractions}. This large difference in mergers may arise from binary-binary encounters, which are not factored in when defining these populations. $a_{\mathrm{ej}}$ as defined in Eq~\ref{eq:aej} assumes that the interaction involves an encounter with a single BH rather than another binary. Since binary-binary encounters are oftentimes more disruptive of a the binaries, if a Pop II BBH undergoes this kind of interaction it may be disrupted\footnote{This could manifest as a "normal" disruption where the target binary is simply broken, or it could result in an \textit{exchange} where the component BHs are swapped, thus "disrupting" the primordial binary.}

In Table~\ref{tab:N-bodyMergers} we also show counts for the number of mergers from the dynamical population. Note that this population includes binaries formed through exchanges at any point in the stellar evolution. We find that these are a minority of mergers compared to the those from the primordial population. The final column of Table~\ref{tab:N-bodyMergers} shows the number of single BHs left in the cluster at the end of the simulation, 1 Gyr. These might still contribute to the dynamical population of merging BBHs.
If we assume that the remaining BHs will interact dynamically to form BBHs and that these BBHs will merge within a Hubble time, then  we can estimate an upper limit for the number of mergers we expect from the dynamical channel. 
This needs to take into account, however, that a binary will eject approximately $\sim 5$ other BHs before merging \citep{breen_dynamical_2013}. We can conclude that the dynamical channel is still expected to produce fewer mergers compared to the primordial channel.

We consider the future evolution of BBH populations we have defined  and conclude that it should be expected for dynamical interactions to ultimately have a larger contribution to the formation of BBHs, since a significant fraction of Pop III binaries will likely experience an exchange at some time. This exchange will likely manifest as a binary-binary interaction and will add to the number of binaries that are formed through captures of the single BHs still in the cluster. The dominance of dynamically formed binaries in these clusters is consistent with previous studies \citep{di_carlo_binary_2020, rastello_dynamics_2021, torniamenti_dynamics_2022} where their N-body simulations showed that the dynamical population was almost always larger than the "original" population \footnote{In these studies, primordial binaries are termed original, whilst dynamical binaries are referred to as "exchange" binaries.}, typically by about a factor of three.

When we look at the merging population we find that the reverse is true, dynamical interactions appear to be less important, with the BBH mergers predominantly arising from the binary population that is mostly unaffected by dynamical interactions. This is also consistent with \citet{di_carlo_binary_2020, rastello_dynamics_2021, torniamenti_dynamics_2022}, who all find that the "original" binaries are more efficient at merging compared to BBHs formed through exchanges. It should be noted that the prevalence of original BBH mergers in our N-body models is even more pronounced which stems from a different time delay distribution of the BBHs at formation (see the solid lines on the left panel of Fig~\ref{fig:Comparetdelay}). When we compare this plot against the same made for the COMPAS results Fig~\ref{fig:GWHub} (re-plotted on the right panel of Fig~\ref{fig:Comparetdelay}), we see that at every metallicity the proportion of BBHs that merge within a Hubble time is larger than found using COMPAS, with the difference more extreme for higher metallicities. This explains why the N-body points shown in Fig~\ref{fig:popfractions} and Fig~\ref{fig:mergefractions} don't match the fractions predicted from COMPAS. It is necessary to check whether this difference in $t_{\mathrm{delay}}$ is arising from the slightly different stellar evolution routines of both codes, or simply from the dynamical environment of the stellar cluster. Therefore, we also complete an "isolated" run in PeTar, with the same initial stellar conditions as the N-body runs but without the dynamics of the stellar cluster. The resulting $t_{\mathrm{delay}}$ distribution is shown in the left panel of Fig~\ref{fig:Comparetdelay}, with the dotted lines showing the distribution for the isolated simulation run. We see that the dynamical environment of the star cluster does impact the $t_{\mathrm{delay}}$ distribution compared to the isolated run, however the isolated run in PeTar is still not able to recover a distribution similar to that found for COMPAS. Therefore, we conclude that the main cause of the discrepancy between the N-body points and the COMPAS lines shown in Fig~\ref{fig:popfractions} and Fig~\ref{fig:mergefractions} must result from the slightly different stellar evolution routines.

\iffalse
\begin{figure}
    \centering
    \includegraphics[width=\columnwidth]{Plots/Comparing_time_delays.pdf}
    \caption{Comparing the time delay distribution across three metallicities for the primordial population found in the N-body results (Solid lines). We only show the results from the $10^{5} \ M_{\sun}$ clusters at each metallicity. We notice that this differs to the same plot made for the COMPAS results Fig~\ref{fig:GWHub}. On here we also show the time delay distribution found using the PeTar code without dynamics (dashed lines). Comparing the dashed and solid lines we can see how much of the difference the dynamical environment of the cluster makes on the BBHs formed.}
    \label{fig:Comparetdelay}
\end{figure}
\fi

\begin{figure*}
    \centering
    \begin{subfigure}{\columnwidth}
        \includegraphics[width=\textwidth]{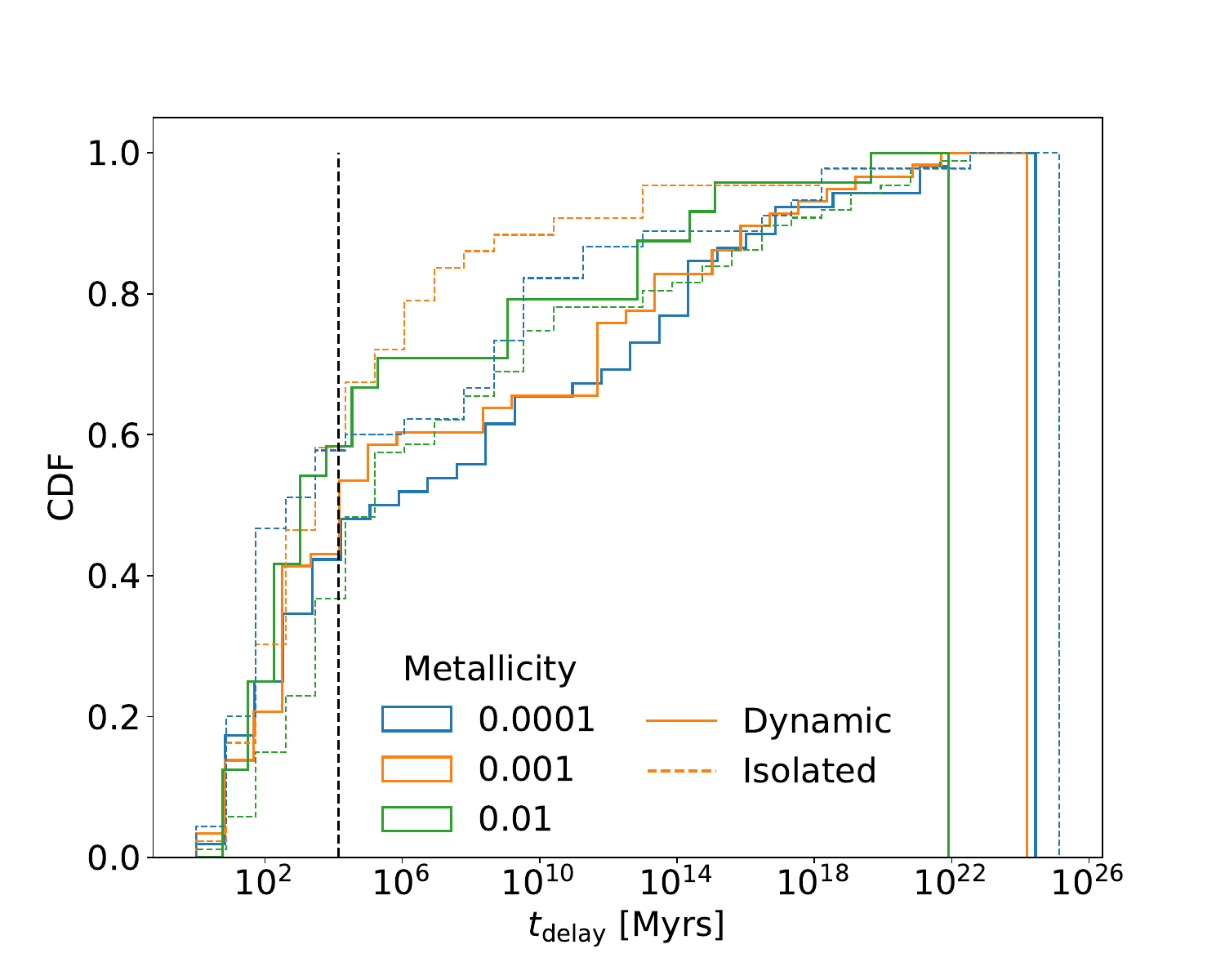}
    \end{subfigure}
    \hfill
    \begin{subfigure}{\columnwidth}
        \includegraphics[width=\textwidth]{Plots/tdelayCDF.pdf}
    \end{subfigure}
    \caption{On the left panel we compare the time delay distribution across three metallicities for the primordial binary population found in the N-body results (Solid lines). We only show the results from the $10^{5} \ M_{\sun}$ clusters at each metallicity. We also show the time delay distribution found using the PeTar code without dynamics (dashed lines). Comparing the dashed and solid lines we can see how much of the difference the dynamical environment of the cluster makes on the BBHs formed. We also note that these distributions differ from the time delay distributions that we previously showed for the COMPAS results (Fig~\ref{fig:GWHub}). We have re-plotted the COMPAS distributions here in the right panel for ease of comparison.}
    \label{fig:Comparetdelay}
\end{figure*}

\begin{table}
    \caption{A count of mergers found within N-body simulations that were ran up to $1 \ \mathrm{Gyr}$. Here the primordial binaries are separated in the three populations based on their orbital parameters at BBH formation. We also give the number of mergers among BBHs that form through dynamical interactions, and the number of lone black holes remaining at the end of the simulation, $N_{\mathrm{BH}}$.}
    \label{tab:N-bodyMergers}
    \begin{tabular}{cccccccc}
        \hline
        Metallicity & Mass & \multicolumn{3}{c}{Primordial} & \multirow{2}{*}{Dynamical} & \multirow{2}{*}{$N_{\mathrm{BH}}$}\\
        $Z_{\sun}$ & $M_{\sun}$ & Pop I & Pop II & Pop III & & \\ \hline
        \multirow{3}{*}{$1Z_{\sun}$} & 10000 & 2 & 0 & 0 & 0 & 1\\
         & 50000 & 4 & 0 & 0 & 0 & 23 \\
         & 100000 & 12 & 2 & 0 & 4 & 50  \\
         \hline
        \multirow{3}{*}{$0.1Z_{\sun}$} & 10000 & 2 & 0 & 0 & 1 & 3\\
         & 50000 & 10 & 1 & 0 & 3 & 29 \\
         & 100000 & 24 & 5 & 1 & 6 & 70 \\
         \hline
        \multirow{3}{*}{$0.01Z_{\sun}$} & 10000 & 4 & 0 & 0 & 1 & 2 \\
         & 50000 & 13 & 1 & 0 & 2 & 34 \\
         & 100000 & 20 & 4 & 1 & 6 & 95 \\ \hline
    \end{tabular}
\end{table}

\subsection{Varying stellar properties}
We vary some of the stellar properties and rerun the analysis performed above, Table~\ref{tab:variations} details all the different models that we run.

\begin{table}
    \caption{Stellar evolution variations compared to the previous models shown above}
    \label{tab:variations}    
    \centering
    \begin{tabular}{c|c}
        \hline
        Model & Variation \\
        \hline
        Mod1 & Standard parameters described above \\
        Mod2 & Chemically homogeneous evolution - Optimistic \\
        Mod3 & Chemically homogeneous evolution - Pessimistic  \\
        Mod4 & BH kick prescription - No kicks \\
        Mod5 & BH kick prescription - Reduced (equal momentum)\\
        \hline
    \end{tabular}
    \label{tab:variations}
\end{table}

We find very little difference between Mod1, Mod2 and Mod3 suggesting that the choice of chemically homogeneous evolution does not affect the population fractions of the BBHs. We also see negligible variation for the BBH mass, semi-major axis, eccentricity and mass ratio distributions across Mod1, Mod2 and Mod3. 

The choice of BH kick prescription has a dramatic effect on the populations and the BH parameters, as can be expected; since the size of the SN kicks is one of the main determinants of whether a binary is a) disrupted following stellar evolution and b) if it remains within the cluster following the BH formation. We investigated three prescriptions for the BH natal kicks; firstly, the "Fallback" model, where the BH kick is scaled by the amount of material that falls back onto the BH. This is what has been used in producing the results up until now. Secondly, we used the "Reduced" kick model, where we assume that the BHs receive the same momentum kick as a typical neutron star, and so the drawn kick magnitude is scaled by the ratio $M_{\mathrm{NS}}/M_{\mathrm{BH}}$. Lastly, we use a zero-kick model where the BHs receive no kicks from the supernovae, although it is still possible for them to experience some kicks due to the mass transfer during the explosion. 

\begin{figure*}
    \centering
    \includegraphics[width=2\columnwidth]{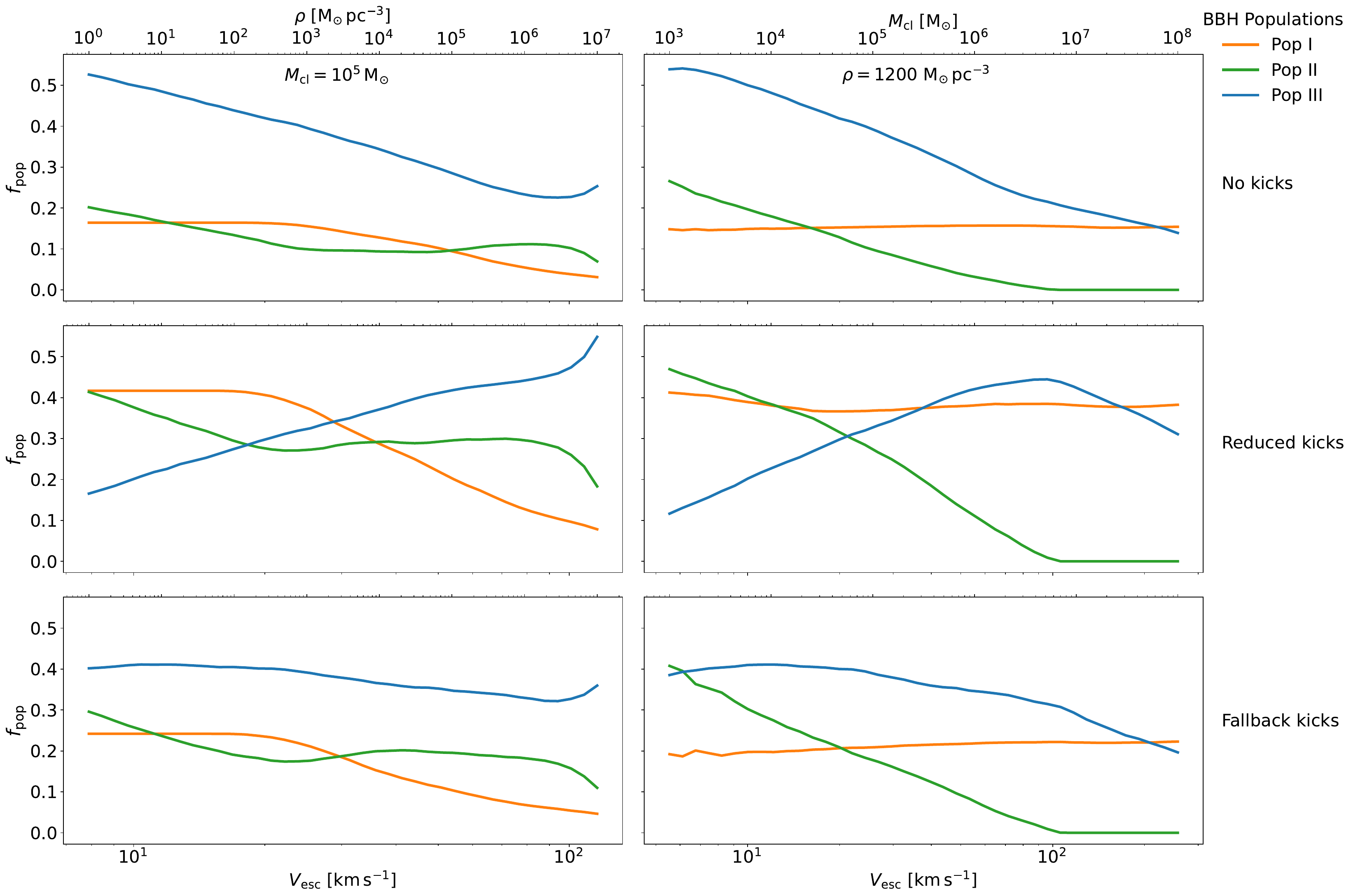}
    \caption{We show the BBH populations, described in caption of Fig~\ref{fig:popfractions}, at metallicity $Z=0.0001\,Z_{\odot}$, assuming different prescriptions for the BH natal kicks. The upper panels show the case for zero natal kicks, the middle panels show the "reduced" model where kicks are scaled by the mass ratio of the BH to a typical NS. Finally, the lower panels show the "fallback" model which scales the kicks based on the amount of material falling onto the BH while it is forming, this is characterised by a fallback fraction $f_{\mathrm{b}}$. These populations are defined as described in the caption of Fig~\ref{fig:popfractions}. The right column of plots assumes a constant density $\rho=1200\,\mathrm{M_{\sun}\,pc}^{-3}$, whereas the left column assumes a constant cluster mass $M_{\mathrm{cl}}=10^{5}\,\mathrm{M_{\sun}}$}.
    \label{fig:BHkicks}
\end{figure*}

We show the population fractions from these results in Fig~\ref{fig:BHkicks} where we have set the metallicity to $Z=0.0001$. When we assume the BHs receive no natal kick (upper panels) we see that the fraction of Pop III BBHs falls rapidly with increasing density and cluster mass which is very different relation to what we had seen when using the fallback models previously (and in the lower panels). Meanwhile relationship between the Pop I fraction and the Pop II fraction with cluster density and mass remain largely the same as in the fallback case, although their respective fractions are slightly lower. This greater dominance in the Pop III BBHs across the escape velocity range is most likely due to the increased number of BBHs that can remain bound following the the BH natal kicks (since these are now zero). Although it is true that the zero kicks should also mean that the number of Pop I and Pop II BBHs should also be increased; Pop III will benefit the most since they have larger separations and thus would be easier to disrupt following a kick.

The reduced kick model (middle panels) is particularly interesting since it has the most apparent change in relationship. In this case we the Pop III BBHs actually start as the minority population and increase in fraction with increasing density and cluster mass. This is contrary to what we see for the fallback kick and zero kick models where the general trend is a decrease in Pop III dominance with escape velocity. This difference is likely arising due to the fact that the reduced prescription for kicks will generally produce larger BH kicks compared to the fallback prescription. These larger kicks are able to disrupt more of the wider Pop III BBHs than the fallback model, and this is reflected in the lower contribution to the BBH population. As the escape velocity increases we have the Pop II/Pop III cut off value, $a_{\mathrm{ej}}$ shrink and so the Pop II BBHs become classified as Pop III, this is the same as what we have discussed previously. However, in this case since we have fewer wider BBHs (those with separations close to the Pop III/soft BBH boundary $a_{\mathrm{h}}$), Pop III gains more members from Pop II than it loses as soft binaries. This explains why we see this increasing trend of dominance in the Pop III with reduced kicks compared to the opposite for the fallback and no kick prescriptions.

\section{Importance of binary-binary interactions}\label{sec:importance}
Thus far we have shown that a significant fraction of the retained BHs are found in BBHs, the latter being the majority source for low $v_{\mathrm{esc}}$ clusters. This abundance of binaries in the cluster will in turn lead to a higher chance that any specific BBH experiences a binary-binary encounter during its lifetime in the cluster. These interactions are complex with a variety of possible outcomes, including exchanges of binary components and complete disruption of one or both binaries \citep{antognini_dynamical_2016,zevin_eccentric_2019}. All of which would naturally have implications on the properties of the BBH population. Therefore, we show the importance of binaries by quantifying the number of potential binary interactions we would expect given the population evolved  with COMPAS, and making some assumptions regarding the host cluster. 

We loosely follow the method shown in \citet{atallah_growing_2022} to calculate the interaction rate between a target binary and a projectile "species", which we  define as the binary or single population as required. Firstly, we set up some relationships for the cluster environment where we are to place our BHs. We assume a double \cite{king_structure_1966} cluster sphere model where  the inner sphere is a scaled-down version of the outer sphere. The inner sphere is taken as the BH sub-cluster which results from the BH mass segregation expected to form in these dense stellar environments \citep{breen_dynamical_2013, kupper_mass_2011}. The outer sphere we take as the primary cluster which we assume to contain only one solar mass stars, such that the average stellar mass for the cluster is $\braket{m_{*}} = 1 \ M_{\sun}$. The two spheres are defined to be uncoupled so that they maintain independent velocity dispersion profiles. Under this assumption, the deviation from energy equipartition between the outer (primary) and inner (BH) cluster at $r=0$ is defined by
\begin{equation}
    \eta = \frac{\braket{m_{\mathrm{BH}}} \sigma_{\mathrm{BH}}^{2}(0)}{\braket{m_{\mathrm{*}}} \sigma_{\mathrm{cl}}^{2}(0)},
    \label{eq:equienergy}
\end{equation}
where $\braket{m_{\mathrm{*}}}$ and $\braket{m_{\mathrm{BH}}}$ are the average masses in the primary and BH clusters. The choice of $\eta$ determines the energy shared between the two clusters, with $\eta=1$ once full equipartition of energy is reached. In this state, the ratio of the velocity dispersion for each cluster scales with the ratio of the average masses $\frac{\sigma_{\mathrm{BH}}^{2}}{\sigma_{\mathrm{*}}^{2}} = \frac{\braket{m_{\mathrm{*}}}}{\braket{m_{\mathrm{BH}}}}$. Alternatively, we can set $\eta=\frac{\braket{m_{\mathrm{BH}}}}{\braket{m_{\mathrm{*}}}}$ which is a state where the BHs have the same velocity dispersion as the stars.

In a similar vein to \citet{atallah_growing_2022}, we assume that the interaction between a target binary  and the projectile species (singles or binaries) occurs within the BH sub-cluster. Then, by assuming equipartition between the target and projectile we can relate their respective velocity dispersions by 
\begin{equation}
    \begin{split}
        \sigma_{\mathrm{p}}  & = \sigma_{\mathrm{BH}}, \\
        \sigma_{\mathrm{t}}  & = \sqrt{\frac{\braket{m_{\mathrm{p}}}}{m_{\mathrm{t}}}} \sigma_{\mathrm{p}}.
    \end{split}
    \label{eq:projveldisp}
\end{equation}
We finally define the relative velocity dispersion in the relative motion frame as in \citet{binney_galactic_2008}.
\begin{equation}
    \sigma_{\mathrm{rel, p}} = \sqrt{\sigma_{\mathrm{t}}^{2} + \sigma_{\mathrm{p}}^{2}} = \sigma_{\mathrm{BH}}\sqrt{1+\frac{\braket{m_{\mathrm{p}}}}{m_{\mathrm{t}}}}.
\end{equation}
We define the escape velocity from the core of both clusters $v_{\mathrm{esc}}(\infty)$, with the total potential at the core $\Phi_{\mathrm{tot}}(0) = \Phi_{\mathrm{cl}}(0)+\Phi_{\mathrm{BH}}(0)$.

\begin{equation}
    v_{\mathrm{esc}}(\infty) = \sqrt{-2(\Phi_{\mathrm{cl}}(0)+\Phi_{\mathrm{BH}}(0))},
\end{equation}
where the potential at infinity goes to 0. From \citet{king_structure_1966} we can relate the central potential to the $W_{\mathrm{0}}$ parameter by $\Phi(0)=-W_{\mathrm{0}}\sigma^{2}$. Setting $W_{\mathrm{0}}=7$ and substituting the velocity dispersion's from Eq~\ref{eq:equienergy} with $\braket{m_{\mathrm{*}}}=1 \ M_{\sun}$ yields

\begin{equation}
    v_{\mathrm{esc}}(\infty) = \sqrt{14\left(1+\frac{\eta}{\braket{m_{\mathrm{BH}}}}\right)}\sigma_{\mathrm{*}}.
\end{equation}

\begin{figure}
    \centering
    \includegraphics[width=\columnwidth]{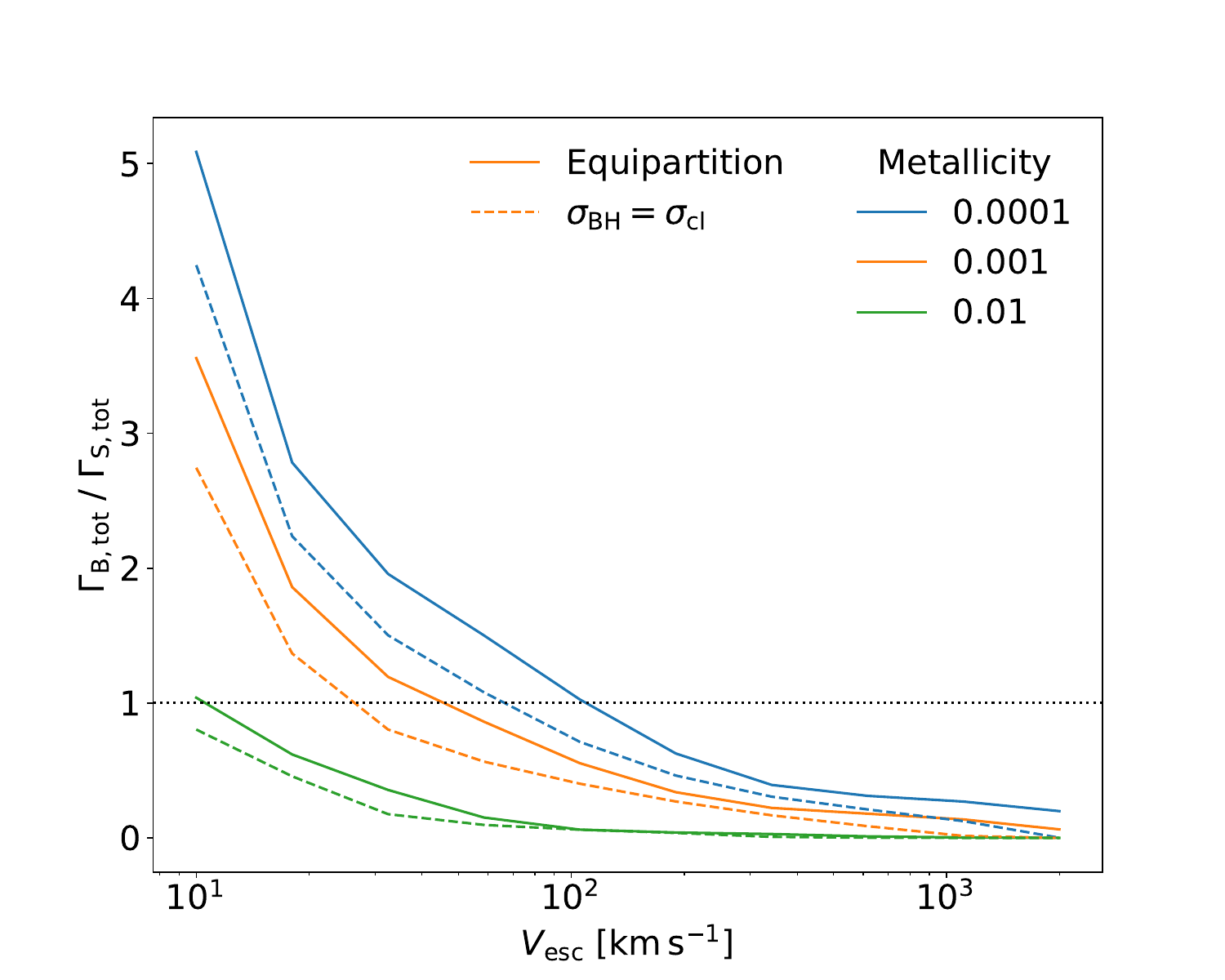}
    \caption{The ratio of the total number of  binary-binary interactions to binary-single interactions, integrated over the entire hard BBH and single BH populations found using COMPAS (see Section~\ref{sec:retain}). We show this ratio across three metallicities ($Z=0.01,\, 0.001,\, \mathrm{and} \ 0.0001$) and for two states of the cluster, energy equipartition and a state of equal velocity dispersion between BHs and stars. We also mark the boundary line of $\Gamma_{\mathrm{B,tot}}/\Gamma_{\mathrm{S,tot}}=1$ below which binary-single interactions become the dominant form of encounter.}
    \label{fig:TotalInt}
\end{figure}

Now we adapt the general interaction rate between a target binary  and a projectile species, such that we integrate the interactions over our hard BBH and BH populations:
\begin{equation}
    \begin{split}
        \Gamma_{\mathrm{s}} & \propto \sum^{N_{\mathrm{s}}}_{i=0} a_{\mathrm{t}}^{2} \sigma_{\mathrm{rel, s}} \left[1 + \frac{G(m_{\mathrm{t}}+m_{\mathrm{s, i}})}{2a_{\mathrm{t}}\sigma^{2}_{\mathrm{rel, s}}} \right] \\
        \Gamma_{\mathrm{b}} & \propto \sum^{N_{\mathrm{b}}}_{i=0} (a_{\mathrm{t}}+a_{\mathrm{b,i}})^{2} \sigma_{\mathrm{rel, b}} \left[1 + \frac{G(m_{\mathrm{t}}+m_{\mathrm{b,i}})}{2(a_{\mathrm{t}}+a_{\mathrm{b, i}})\sigma^{2}_{\mathrm{rel, b}}} \right] \ ,
    \end{split}
    \label{eq:interactionRates}
\end{equation}
Where $a_{\mathrm{t}}$ and $a_{\mathrm{b,i}}$ are the target and projectile binary separations and $\sigma_{\mathrm{rel,b}}$ and $\sigma_{\mathrm{rel, s}}$ is the relative velocity dispersion assuming binary and single projectiles respectively. We can further integrate over the target binary semi-major axis and mass, which simply includes a second summation in Eq~\ref{eq:interactionRates} over $a_{\mathrm{t, j}}$ and $m_{\mathrm{t, j}}$. In doing this we can estimate the total interaction rates for every BBH in the population with every other BBH and with every single BH in the population. The ratio of these two rates can give us a measure of the dominant form of interactions given our BH populations. Fig~\ref{fig:TotalInt} shows the ratio of the total interaction rate for every binary-binary ($\Gamma_{\mathrm{b, tot}}$) and binary-single ($\Gamma_{\mathrm{s, tot}}$) encounter in our population against the cluster escape velocity. We plot this curve at three metallicities $Z=0.01, \ 0.001, \ \mathrm{and} \ 0.0001$ and distinguish the boundary line at $\Gamma_{\mathrm{b, tot}}/\Gamma_{\mathrm{s, tot}} = 1$ below which single interactions become more dominant in the population. In addition, we consider two possible states of the cluster; a state of energy equipartition where $\frac{\sigma_{\mathrm{BH}}^{2}}{\sigma_{\mathrm{*}}^{2}} = \frac{\braket{m_{\mathrm{*}}}}{\braket{m_{\mathrm{BH}}}}$ and a state where the velocity dispersion of the BHs and stars is equal. One would expect that as a stellar cluster evolves, it moves towards a state of energy equipartition. 

Concentrating first on the case where the cluster has reached energy equipartition; the trivial takeaway from Fig~\ref{fig:TotalInt} is that binary-binary interactions dominate the BBH encounters at low $v_{\mathrm{esc}}$. 
%This makes physical sense since we have shown previously that it is in the smaller clusters where the BH population is predominantly found within BBHs as the single BHs typically are given SN kicks large enough to escape the cluster entirely.
As the escape velocity increases and we begin to retain more single BHs within the cluster the number of potential binary-single encounters grows. In addition, the velocity dispersion of the cluster increases and as such the hard-soft boundary for the BBHs ($a_{\mathrm{h}}$) decreases and so does the number of hard binaries which can undergo a binary-binary encounter. The combination of these two effects leads to the ratio of $\Gamma_{\mathrm{B, tot}}/\Gamma_{\mathrm{S, tot}}$ decreasing as the escape velocity increases.
%, to such a degree where we transition from a cluster where binary-binary interactions dominate, to a cluster where binary-single interactions dominate. 
We see that the escape velocity where $\Gamma_{\mathrm{B, tot}}/\Gamma_{\mathrm{S, tot}}$ becomes less than one is dependent on the metallicity of the cluster, with the lower metallicity clusters maintaining a majority of binary-binary interactions for higher escape velocities. We also see that the escape velocity of the transition point, $\sim 100$ \kms for the sub-Solar metallicity model, is larger than the escape velocity at which the single BH population becomes more numerous than the hard BBH population, $~30$ \kms for the same model. This tells us that this  transition point is not only dependent on the number count of single BHs to BBHs. For each interaction, we have a cross-section for the interaction $\Sigma_{\mathrm{int}} \propto k(a_{\mathrm{t}}+a_{\mathrm{p}})^{2}$ with $a_{\mathrm{p}}=0$ when the projectile species are single BHs and $k=2$ so that we include only strong interactions with the target binary. Naturally, the interaction cross section for a binary-binary interaction is larger than for a binary-single interaction. Thus, for the binary-single interactions to become dominant, they not only need to outnumber the binary-binary interactions but outnumber them to such a degree that the extra encounters can make up for their lower cross-section. To better understand this cut-off, we investigated how the binary fraction, $N_{\mathrm{bin}}/(N_{\mathrm{bin}}+N_{\mathrm{singles}})$, evolves with $v_{\mathrm{esc}}$. We find that for every metallicity model the binary-single interactions become dominant below a binary fraction of $30\,\%$, which is consistent with recent work on the topic \citep{marin_pina_dynamical_2023}.

For each metallicity model in Fig~\ref{fig:TotalInt} we also consider the case where $\sigma_{\rm BH}=\sigma_{\rm *}$. We see that in this scenario the transition point to dominant binary-single interactions is pushed to lower escape velocities for all models, while the general shape of the relationship between $\Gamma_{\mathrm{B, tot}}/\Gamma_{\mathrm{S, tot}}$ and $v_{\mathrm{esc}}$ remains almost unchanged compared to the equipartition case. 
%The reason for this is simple, the velocity dispersion for the BHs and BBHs, in this case, is larger than in the equipartition case which in turn leads to the hard-soft boundary $a_{\mathrm{h}}$ for the BBHs being reduced by a factor of $1/\braket{M_{\mathrm{BH}}}$. Naturally, this means that at any given escape velocity there now exists fewer hard binaries to be considered for an encounter compared to the same escape velocity in the equipartition case, this has the effect of simply shifting the relationships to smaller $v_{\mathrm{esc}}$

\begin{figure}
    \centering
    \includegraphics[width=\columnwidth]{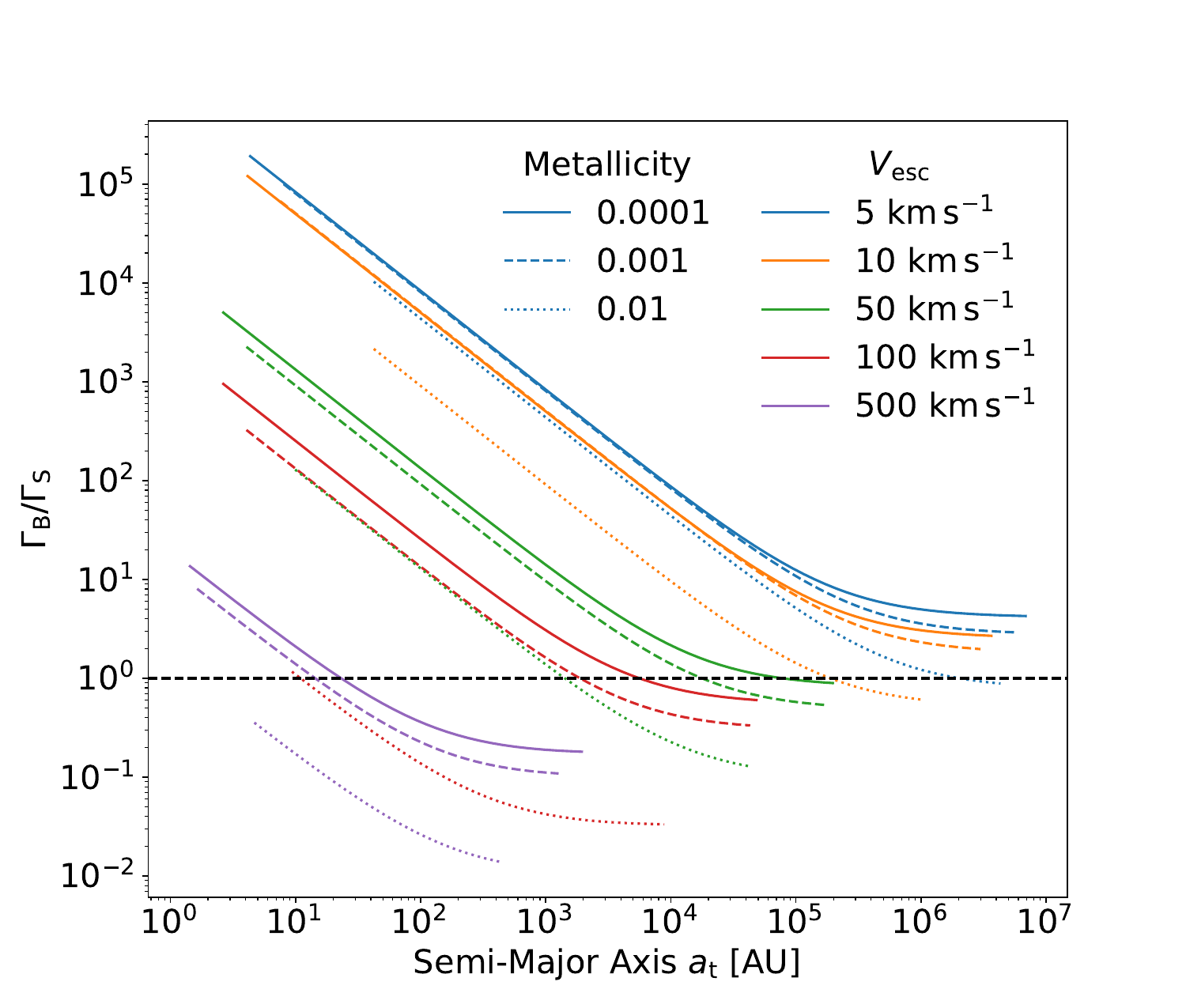}
    \caption{For our three metallicity models $Z=0.01,\, 0.001,\, 0.0001$ we assume five cluster escape velocities, $v_{\mathrm{esc}} = 5,\, 10,\, 50,\,100,\, \mathrm{and}\, 500$ \kms and then generate 10,000 target binaries drawing their separations from the range of semi-major axis in the retained hard BBH population. We then calculate the ratio of binary-binary and binary-single interaction rates for each target binary, integrating  over the retained hard BBH and single BH populations. We finally show this ratio against the target semi-major axis for all three metallicities and all five escape velocities, in addition to plotting the boundary line at $\Gamma_{\mathrm{B}}/\Gamma_{\mathrm{S}}=1$ below which the binary-single interactions are dominant}
    \label{fig:semiInt}
\end{figure}

To investigate how the ratio $\Gamma_{\mathrm{B, tot}}/\Gamma_{\mathrm{S, tot}}$ depends on the properties of the target binary, we run further analysis and calculate the interaction rates for a given target BBH with our hard BBH and single BH populations. %Since the interaction rates given by Eq~\ref{eq:interactionRates} are only ever dependent on the mass and the semi-major axis of the target binary, these are the only two parameters of the target binary that we need to vary.
Starting first by fixing the target BBH mass to the average BBH mass from our population, $\braket{m_{\mathrm{BBH}}}=44 M_{\sun}$, and fixing the escape velocity, we draw 10,000 target separations uniformly across the range of hard BBH semi-major axis in our population. For each target binary, we calculate the averaged ratio of the interactions $\Gamma_{\mathrm{B}}/\Gamma_{\mathrm{S}}$, repeating this for five different values of $v_{\mathrm{esc}}$ and also for each of our metallicity models. Fig~\ref{fig:semiInt} shows these results plotted against the semi-major axis of the target binary ($a_{\mathrm{t}}$), again with the cut-off value $\Gamma_{\mathrm{B}}/\Gamma_{\mathrm{S}}=1$ marked. 

We see that for a given metallicity and escape velocity, the ratio of interactions increases for a smaller target separation, with the relationship at low $a_{\mathrm{t}}$ scaling proportional to $~\left(1+\frac{\braket{a_{\mathrm{b}}}}{a_{\mathrm{t}}}\right)$. Looking to the other extreme of  large  $a_{\mathrm{t}}$,  at all escape velocities and metallicities the ratio of interactions levels out. It can be shown that in this regime 
\begin{equation}
    \frac{\Gamma_{\mathrm{B}}}{\Gamma_{\mathrm{S}}} \simeq \frac{N_{\mathrm{B}}}{N_{\mathrm{S}}}\frac{\sigma_{\mathrm{rel, B}}}{\sigma_{\mathrm{rel, S}}}  \ ,
    \label{eq:higha}
\end{equation}
where we assume that the target separation is much larger than the average separation of the projectile binaries. Note that since here we are fixing the escape velocity of the cluster, Eq~\ref{eq:higha} is constant and so in the high $a_{\mathrm{t}}$ regime the key factor determining whether binary-single or binary-binary interactions are dominant comes down to which population is more numerous at this $v_{\mathrm{esc}}$. This explains why for low escape velocities, $v_{\mathrm{esc}} = 5$ \kms and $v_{\mathrm{esc}}=10$ \kms where we are retaining many more hard BBHs than single BHs, the single interactions are never dominant, even at large $a_{\mathrm{t}}$.

As the cluster escape velocity increases, and so, in turn, does the velocity dispersion, we know that $a_{\mathrm{h}}$ for any given binary decreases which both limits the maximum separation of the target binary and lowers the number of hard BBH to be used as projectiles. In addition, the cluster retains more single BHs at higher escape velocity which naturally increases the binary-single interaction rate of the target binary. All of these aspects can be seen in Fig~\ref{fig:semiInt}, when comparing curves at increasing $v_{\mathrm{esc}}$. The key point here is that for a given target binary, the separation at which the binary-single interactions become dominant becomes smaller as the escape velocity of the cluster increases. In particular, for a metallicity $Z=0.0001$, binary-binary interactions are the dominant form of encounter in clusters up to $v_{\mathrm{esc}}=100$ \kms for almost any target binary separation.
%which is a result of both the greater retention of single BHs at higher escape velocities, the reduced number of hard BBHs which also have smaller average separation leading to smaller average cross-section for the interaction.

\begin{figure}
    \centering
    \begin{subfigure}{\columnwidth}
        \includegraphics[width=\columnwidth]{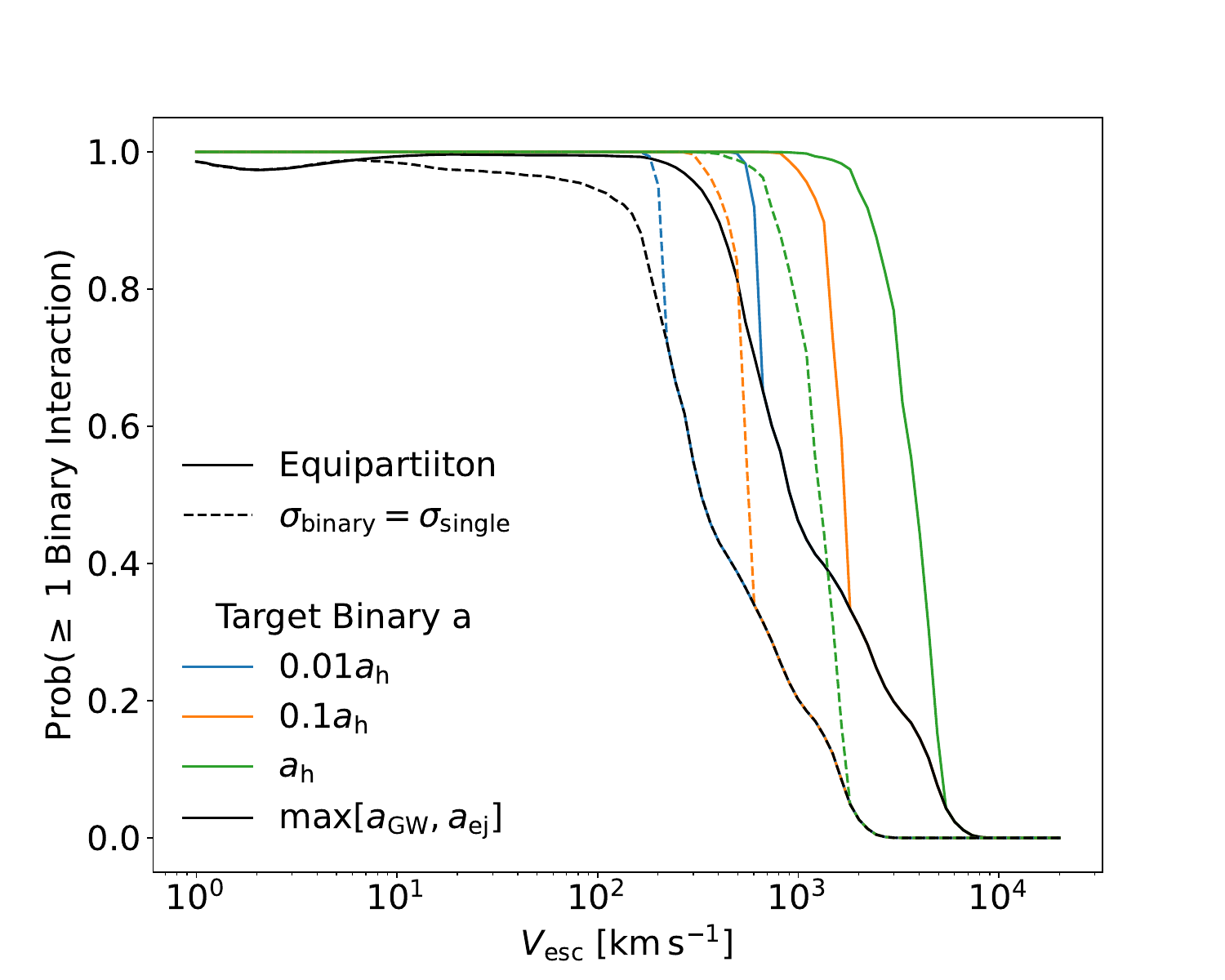}    
    \end{subfigure}
    \begin{subfigure}{\columnwidth}
        \includegraphics[width=\columnwidth]{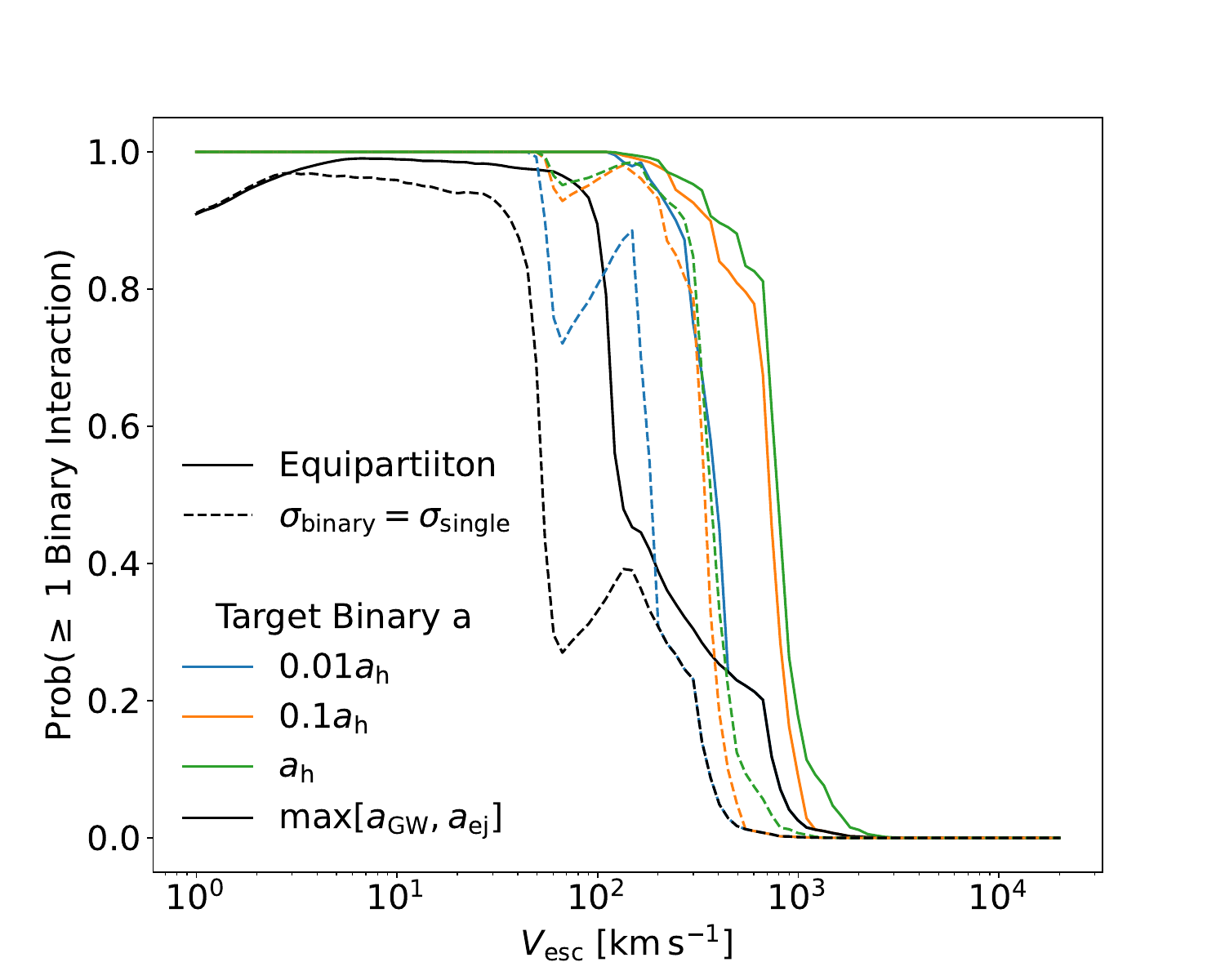}    
    \end{subfigure}
    
    \caption{Probability to experience at least one binary-binary encounter in the chain of successive interactions a target binary undergoes before its separation reduces to $a_{\rm ej}$. The probability is computed as 1-$P(\mathrm{All~Singles})$, where $P(\mathrm{All~Singles})$ is the probability that the binary only interacts with singles. We show this probability by taking three values for the initial separation of the target binary in terms of the hard/soft boundary, $a_{\mathrm{h}}$, $0.1a_{\mathrm{h}}$ and $0.01a_{\mathrm{h}}$. We also compute the probability of the next encounter being a binary for a target binary with separation $a=\max[a_{\mathrm{GW}}, \ a_{\mathrm{ej}}]$ where $a_{\mathrm{GW}}$ is the separation at which the binary energy loss is dominated by GW radiation. Finally, we perform the above analysis assuming two states for the overall cluster: a state of energy equipartition, and a state with the BH velocity dispersion equal to the stellar velocity dispersion. The upper panel shows results for $Z=0.0001$ while the lower panel shows for $Z=0.01$.}   
    \label{fig:ProbInteraction}
\end{figure}

\subsection{Probability of at least one binary-binary encounter}
Whilst within the cluster, a hard BBH will undergo multiple encounters until it is either disrupted or ejected from the cluster. Although interactions with other binaries can cause some chaotic outcomes, it is typical to assume strong encounters with single BHs will result in the binary giving up about 20\% of its energy, thus shrinking its semi-major axis. This tightening of the binary will continue with successive encounters until its separation reaches $a_{\mathrm{ej}}$ as defined in Eq~\ref{eq:aej} at which point the next single interaction will eject the binary from the cluster \citep{antonini_merging_2016}. However, when the cluster starts with a non-zero binary fraction, we can expect that there are many other BBHs (see Fig~\ref{fig:retainedBHsnew}) in the cluster which will complicate the simple picture above, since at any point along this chain of interactions, the target binary may interact with another binary. Strong binary-binary interactions will typically result in a greater change in the binding energy of one of the binaries, compared to a typical binary-single encounter. In addition to the increased effect of hardening, binary-binary interactions can also lead to exchanges of the binary components which results in essentially a new, dynamically formed, binary \citep{zevin_eccentric_2019}. %When this happens it is not clear that the target binary actually survives the encounter. 

Given our populations of BHs and BBHs formed solely from binary stellar evolution, we calculate the probability that a target binary interacts with at least one other binary before it can shrink its separation to $a_{\mathrm{ej}}$ (Eq~\ref{eq:prob}). Here $P_{\mathrm{int}}(\mathrm{All\,Singles})$ is the probability that the target binary receives only $N$ single interactions, where $N=\log{\left(\frac{a_{\mathrm{h}}}{a_{\mathrm{ej}}}\right)}/\log{1.2}$ is the number of single interactions to reach $a_{\mathrm{ej}}$ from the initial separation $a_{\mathrm{init}}$
\begin{equation}
    P_{\mathrm{int}}(\mathrm{Binary}) = 1 - P_{\mathrm{int}}(\mathrm{All \ Singles}) .
    \label{eq:prob}
\end{equation}

Using the interaction rates in Eq~\ref{eq:interactionRates} we can define the probability for the next interaction to be with a single BH, given the current separation of the target binary 
\begin{equation}
    P_{\mathrm{int}}(\mathrm{single}) = \frac{\Gamma_{\mathrm{S}}}{\Gamma_{\mathrm{S}}+\Gamma_{\mathrm{B}}}.
\end{equation}
We then calculate this $P_{\mathrm{int}}(\mathrm{single})$ $N$ times as the target binary's separation shrinks to $a_{\mathrm{ej}}$. The product of these probabilities then gives us the probability of only interacting with single BHs, $P_{\mathrm{int}}(\mathrm{All ~Singles})$, which we put into Eq~\ref{eq:prob} to find the probability of at least one binary interaction. We perform this calculation for three distinct target binaries, all with $M_{\mathrm{tot}}=44 \ M_{\sun}$, and separations as fractions of the hard/soft boundary, $0.01a_{\mathrm{h}}, \ 0.1a_{\mathrm{h}}$ and $a_{\mathrm{h}}$. Fig~\ref{fig:ProbInteraction} takes the data from the $Z=0.0001$ model and plots this probability for a range of cluster escape velocities, also taking the two possible assumptions about the cluster state, equipartition and equal velocity dispersion. We see that for low $v_{\mathrm{esc}}$ clusters the probability of encountering a binary before being ejected is $\approx 1$, while as the escape velocity increases this probability drops off quite rapidly. %This is not somewhat unsurprising since we showed previously that the smaller clusters at this metallicity are dominated by binary-binary encounters. 
However, the probability of at least one binary encounter for a target binary initially at $a_{\mathrm{h}}$ remains almost certain up to a $v_{\mathrm{esc}}\simeq10^{3}$ \kms. Calculating the number of single interactions required for a binary to shrink from $a_{\mathrm{h}}$ to $a_{\mathrm{ej}}$, we find that it takes $\sim30$ encounters. Since $a_{\mathrm{h}}$ and $a_{\mathrm{ej}}$ both scale with $1/v_{\mathrm{esc}}^{2}$, this number of encounters is independent of the cluster. Thus, even at the highest escape velocities, it would still take 30 encounters to reach $a_{\mathrm{ej}}$ and so even if every interaction in that chain had a 90\% chance of being with a single BH, the probability of all 30 being encounters involving a single BH is still only 4\%. 
%All this means that for a BBH to never interact with another binary before being ejected, you need a very high escape velocity cluster where $\Gamma_{\mathrm{B}}/\Gamma_{\mathrm{S}}<<1$, so that the probability of a single interaction approaches 100\%.

When the cluster escape velocity increases, the target binary enters a regime with $a_{\mathrm{ej}}<a_{\mathrm{GW}}$, where $a_{\mathrm{GW}}$ is defined as the separation at which the gravitational wave radiation dominates the energy loss of the binary  \citep{antonini_merging_2016}
\begin{multline}
  a_{\mathrm{GW}}=0.05 \left(\frac{M_{\mathrm{tot}}}{20 \ M_{\sun}}\right)^{3/5} \left(\frac{q}{(1+q)^{2}}\right)^{1/5} \\ \left(\frac{\sigma_{\mathrm{rel}}}{30 \ \mathrm{km s}^{-1}}\right)^{1/5}\left(\frac{10^{6} \ M_{\sun}\mathrm{pc}^{-3}}{\rho}\right)^{1/5} \ \mathrm{AU} 
  \ . \label{eq:aGW}
\end{multline}  
Once in this regime, the number of single interactions that the binary experiences start to decrease with increasing $v_{\mathrm{esc}}$. Since $a_{\mathrm{GW}}$ is only weakly dependant on cluster properties $\sigma_{\mathrm{rel}}$ and $\rho$ compared to the binary hard/soft boundary and so $a_{\mathrm{h}}$ rapidly approaches $a_{\mathrm{GW}}$ as $v_{\mathrm{esc}}$ increases. The BBH now needs to experience fewer single encounters before it eventually reaches $a_{\rm{GW}}$ and then merges before another interaction. Thus the total probability that at least one of the interactions is with another binary decreases. We see this at the high $v_{\mathrm{esc}}$ regime of Fig~\ref{fig:ProbInteraction}, with the probability of each target binary eventually decreasing.
In Fig~\ref{fig:ProbInteraction} we have also calculated the probability that the next encounter would be with a binary if the target binary has a separation $a=\max(a_{\rm{GW}}, \ a_{\mathrm{ej}})$.

\section{Conclusions}
\label{sec:final}
In this work, we have used the  binary population synthesis code, COMPAS \citep{riley_rapid_2022, stevenson_formation_2017} to characterise the population of BBHs that form by stellar processes in star clusters. Unlike dynamically formed BBHs, the  properties (e.g., component masses, orbit) of such  primordial binaries are set mostly by  stellar evolution processes. After their formation, however, they can  undergo dynamical interactions that change their orbit and their likelihood of becoming a detectable source of GW radiation. These binaries represent therefore a hybrid population, in the sense  that they can be significantly  affected by both stellar and dynamical processes. 

   We have presented  simple analytical arguments together with binary evolution models and $N$-body simulations to study the formation and evolution of primordial  BBHs in dense star clusters. These models represent a baseline for understanding their contribution  to the population of merging BBHs detectable by LIGO-Virgo-Kagra.      
We  briefly investigate how the choice of stellar ZAMS metallicity affects the binary properties of the BBHs that are formed. We then focus our efforts on investigating the effect of placing the BBH and single BH populations in simplistic cluster models. We compare the kicks received during the SN; as well as the expected kicks due to many-body interactions, against a range of cluster escape velocities. From this, we estimate the fraction of BBHs and single BHs that could be retained within different-sized clusters, as well as the sub-population of merging BBHs both inside and outside the cluster. Finally, we study the type of interactions these binaries are likely to experience when evolving in the dynamical environment of their parent cluster.
The key conclusions we find are as follows:

\begin{itemize}
    \item {In  clusters with escape velocity $v_{\mathrm{esc}}\lesssim 100$\kms, BHs are predominantly found as BBH, with a significant fraction also being categorised as "hard" binaries. We expect therefore that primordial binaries might have a significant impact on the merger rate of BBHs formed in open and globular clusters. On the other hand, we expect  their contribution to be smaller in higher velocity dispersion clusters such as nuclear star clusters. }

    \item {The retained BBH population can be further split into three distinct groups based on the binary separation compared to the ejection separation $a_{\mathrm{ej}}$, defined as the separation below which a dynamical interaction will eject the binary from the cluster.
    
    \begin{itemize}
        \item {Pop I: These are binaries with $a<a_{\mathrm{ej}}$ that are so tightly bound that they either merge inside the cluster before an encounter can interfere with them, or they are ejected by the SN kick and merge outside the cluster.}
        \item {Pop II: These binaries also have $a\leq a_{\mathrm{ej}}$, however, they will experience a single interaction which ejects them from the cluster. This group may also merge in a Hubble time though is not defined to do so.}
        \item {Pop III: The final population are hard binaries with $a>a_{\mathrm{ej}}$ and will experience more than one interaction inside the cluster. This group has the most uncertain future as it could eventually be disrupted, become ejected or even merge.}
    \end{itemize}
    }

    \item {When further constraining Pop II and Pop III to those potentially merging in a Hubble time, we see that mergers in a cluster with $v_{\mathrm{esc}}\lesssim 100$ \kms are predominantly Pop I and Pop II. Meanwhile, the Pop III mergers become dominant in the higher $v_{\mathrm{esc}}$ clusters.}

    \item {When using an N-body simulation code to evolve realistic cluster models, we find that the Pop I mergers are dominant with respect to the other two populations. Our models suggest that in clusters with escape velocity $v_{\mathrm{esc}}\leq30$ \kms  dynamics play a secondary role in the production of BBH mergers.
    %We also find that for small clusters $v_{\mathrm{esc}}\leq30$ \kms the dynamically formed binaries can only produce at most twice the number of mergers from the primordial population. This number goes down when we also consider that each dynamically formed binary will, on average, eject 10 BHs. 
    }
    
    \item {Interactions within the cluster are dominated by binary-binary encounters for cluster sizes up to $v_{\mathrm{esc}}\lesssim 100$ \kms for $Z=0.0001$, up to $v_{\mathrm{esc}}\lesssim 40$ \kms for $Z=0.001$, and  up to $v_{\mathrm{esc}}\lesssim 10$ \kms for Solar metallicity. 
    This is of particular importance to Pop III BBHs which experience multiple interactions. For these, it becomes almost certain that at least one of the interactions they experience will be with another hard BBH.}
\end{itemize}

In addition, we tested models with varied stellar evolution parameters to investigate how these new populations are impacted by the choice of evolution. Of particular note, we varied the BH natal kick prescription between a zero kick, fallback and reduced kick model. We find that in the zero kick case, Pop III becomes the dominant group down to $v_{\mathrm{esc}}=4$ \kms. Whereas, in both the fallback and reduced kick models, Pop II become more significant at the small $v_{\mathrm{esc}}$ regime, with Pop III only rising to dominance at $v_{\mathrm{esc}}=80$ \kms with the reduced kicks, and $v_{\mathrm{esc}}=14$ \kms for the fallback model.

Our results indicate that primordial binaries can have a significant impact on the population of merging BBHs produced in dense star clusters.   The initial BH population in  clusters with sufficiently low escape velocities, $v_{\mathrm{esc}}\lesssim 30$ \kms, can be entirely in the form of hard BBHs that originate by stellar processes. The properties of the merging BBHs produced in these clusters are expected to be determined mostly by stellar evolution, with little or no effect from dynamics. 
%   This also implies that evolution of BBHs inside the cluster is dominated by binary-binary interactions as opposed to binary-single interactions, contrary to what is sometimes assumed in the literature \citep[e.g.,][]{samsing_eccentric_2018,antonini_merger_2020,mapelli_hierarchical_2021,arca_sedda_isolated_2023}. 
An implication is that the enhancement of the merger rate due to dynamics is expected to be negligible in these systems.
In clusters with  higher escape velocities, the primordial binary population becomes progressively less important. However, in the range $v_{\mathrm{esc}}\lesssim 100\,$\kms, more than $10\%$ of all BHs are still in hard binaries. A large fraction of these are so tight that they are ejected from the cluster after one dynamical interaction and then merge in the field. This population is of particular interest as their binary properties are set mostly by stellar evolution, but can include some influence due to the single interaction that ejects them. Finally, in higher escape velocity clusters, the single binary population becomes dominant as most of the binaries formed by stellar processes are soft and quickly disrupted.

\section*{Acknowledgements}
Simulations in this paper utilised the COMPAS rapid population synthesis code version (v02.33.02) which is freely available at \url{https://github.com/TeamCOMPAS/COMPAS}. The N-body models utilised the PeTar high-performance N-body code (version 1047\_290) which is free available at \url{https://github.com/lwang-astro/PeTar}. We acknowledge the support of the Supercomputing Wales project, which is part-funded by the European Regional Development Fund (ERDF) via Welsh Government. JB is supported by the STFC grant ST/T50600X/1, DC is supported by the STFC grant ST/V005618/1, and FA is supported by an STFC Rutherford fellowship (ST/P00492X/2)

%%%%%%%%%%%%%%%%%%%%%%%%%%%%%%%%%%%%%%%%%%%%%%%%%%
\section*{Data Availability}
The data used for this work will be freely shared upon reasonable request to the author.



\typeout{}

\bibliographystyle{mnras}
\bibliography{biblio} % if your bibtex file is called example.bib

%\begin{thebibliography}{99}
%\end{thebibliography}

%%%%%%%%%%%%%%%%%%%%%%%%%%%%%%%%%%%%%%%%%%%%%%%%%%

%%%%%%%%%%%%%%%%% APPENDICES %%%%%%%%%%%%%%%%%%%%%

\appendix

%%%%%%%%%%%%%%%%%%%%%%%%%%%%%%%%%%%%%%%%%%%%%%%%%%

% Don't change these lines
\bsp	% typesetting comment
\label{lastpage}
\end{document}